%% file: main.tex
\documentclass[journal, 9pt]{IEEEtran}

\def\supplementary{1}

\ifCLASSINFOpdf
   \usepackage[pdftex]{graphicx}
  \graphicspath{{../pdf/}{../jpeg/}}
  \DeclareGraphicsExtensions{.pdf,.jpeg,.png}
\else
  \usepackage[dvips]{graphicx}
  \graphicspath{{../eps/}}
  \DeclareGraphicsExtensions{.eps}
\fi

\usepackage[cmex10]{amsmath}
\usepackage{amsthm}
\usepackage{amsfonts}
\usepackage{MnSymbol}
\usepackage{stfloats}
\usepackage{tikz}

\usepackage{graphicx}
\usepackage{array}
\usepackage{multirow}
\usepackage{hhline}
\usepackage{makecell}
\usepackage{pbox}
\usepackage[noadjust]{cite}
\usepackage{subfigure}
\usepackage{color}
\usepackage[export]{adjustbox}

\input{macros}

\usepackage{hyperref}
\usepackage{url}

\begin{document}
\title{DCT and DST Filtering with Sparse Graph Operators}
\author{Keng-Shih~Lu,~\IEEEmembership{Member,~IEEE}, Antonio~Ortega,~\IEEEmembership{Fellow,~IEEE}, Debargha~Mukherjee,~\IEEEmembership{Senior~Member,~IEEE}, and~Yue~Chen%
\thanks{K.-S.~Lu, D.~Mukherjee and Y.~Chen are with Google, Mountain View, CA 94043, USA (email: kslu@google.com; debargha@google.com; yuec@google.com). Most of this work has been done while K.-S.~Lu was a PhD student at USC.}
\thanks{A.~Ortega is with the Department
of Electrical and Computer Engineering, University of Southern California, Los Angeles, CA 90089, USA (email: ortega@sipi.usc.edu).}}%



\maketitle

\begin{abstract}
Graph filtering is a fundamental tool in graph signal processing.
Polynomial graph filters (PGFs), defined as polynomials of a fundamental graph operator, can be implemented in the vertex domain, and usually have a lower complexity than frequency domain filter implementations. 
In this paper, we focus on the design of filters for graphs with graph Fourier transform (GFT) corresponding to a discrete trigonometric transform (DTT), i.e., one of 8 types of discrete cosine transforms (DCT) and 8 discrete sine transforms (DST). In this case, we show that multiple sparse graph operators can be identified, which allows us to propose a generalization of PGF design: \emph{multivariate polynomial graph filter (MPGF)}. First, for the widely used DCT-II (type-2 DCT), we characterize a set of sparse graph operators that share the DCT-II matrix as their common eigenvector matrix. This set contains the well-known connected line graph. These sparse operators can be viewed as graph filters operating in the DCT domain, which allows us to approximate any DCT graph filter by a MPGF, leading to a design with more degrees of freedom than the conventional PGF approach. 
Then, we extend those results to all of the 16 DTTs as well as their 2D versions, and show how their associated sets of multiple graph operators can be determined. 
We demonstrate experimentally that ideal low-pass and exponential DCT/DST filters can be approximated with higher accuracy with similar runtime complexity. Finally, we apply our method to transform-type selection in a video codec, AV1, where we demonstrate significant encoding time savings, with a negligible compression loss. 
\end{abstract}


\IEEEpeerreviewmaketitle

\section{Introduction}
\label{sec:introduction}

Graph signal processing (GSP) \cite{sandryhaila2013discrete,shuman2013emerging,ortega2018graph} extends classical signal processing concepts to data living on irregular domains. In GSP, the data domain is represented by a graph, and the measured data is called graph signal, where each signal sample corresponds to a graph vertex, and relations between samples are captured by the graph edges. 
%
Filtering, where frequency components of a signal are attenuated or amplified, is a fundamental operation in signal processing.  
Similar to conventional filters in digital signal processing, which manipulate signals in Fourier domain, a graph filter 
can be characterized by a frequency response that indicates how much the filter amplifies each graph frequency component. This notion of frequency selection leads to various applications, including graph signal denoising \cite{chen2014signal,onuki2016graph,yagan2016spectral}, classification \cite{ma2016diffusion} and clustering \cite{tremblay2016compressive}, and graph convolutional neural networks \cite{kipf2017semi-supervised,defferrard2016convolutional}.

For an undirected graph, a frequency domain graph filter operation $\yv=\Hm\xv$ with input signal $\xv$ and filter matrix
\begin{equation}
\label{eq:graph_filter_freq}
\Hm=\Phim \cdot h(\Lambdam) \cdot \Phim^\top, \quad h(\Lambdam):=\diag(h(\lambda_1),\cdots,h(\lambda_N))
\end{equation}
involves a forward graph Fourier transform (GFT) $\Phim^\top$, a frequency selective scaling operation $h(\Lambdam)$, and an inverse GFT $\Phim$. However, as fast GFT algorithms are only known for graphs with certain structural properties \cite{lu2019fast}, the GFT can introduce a high computational overhead when the graph is arbitrary. To address this issue, graph filters can be implemented with polynomial operations in vertex domain:
\begin{equation}
\label{eq:pgf}
  \Hm = \sum_{k=0}^K g_k \Zm^k, \quad\text{with } \Zm^0=\Id,
\end{equation}
where the $g_k$'s are coefficients and $\Zm$ is called the \emph{graph shift operator}, \emph{fundamental graph operator}, or \emph{graph operator} for short. With this expression, graph filtering can be applied in the vertex (sample) domain via $\yv=\Hm\xv$, which does not require GFT computations. 
A graph filter in the form of \eqref{eq:pgf} is usually called an \emph{FIR graph filter} \cite{isufi2017autoregressive,coutino2019advances} as it can be viewed as an analogy to conventional FIR filters with order $K$, which are polynomials of the delay $z$. In this paper, we call the filters defined in \eqref{eq:pgf} \emph{polynomial graph filters (PGFs)}.

Various methods for designing vertex domain graph filters given a desired frequency response have been studied in the literature. Least squares design of polynomial filters given a target frequency response was introduced in \cite{tremblay2018design,sandryhaila2014discrete,ortega2021introduction}. The recurrence relations of Chebyshev polynomials provide computational benefits in distributed filter implementations
as shown in \cite{hammond2011wavelets,shuman2018chebyshev}. 
In \cite{segarra2017optimal} an extension of graph filter operations to a node-variant setting is proposed, along with polynomial approximation approaches using convex optimization. Autoregressive moving average (ARMA) graph filters, whose frequency responses are characterized by rational polynomials, have been investigated in \cite{loukas2015distributed,isufi2017autoregressive} in both static and time-varying settings. Design strategies of ARMA filters are further studied in \cite{liu2019filter}, which provides comparisons to PGFs. Furthermore, in \cite{coutino2019advances}, state-of-the-art filtering methods have been extended to an edge-variant setting. 
All these methods are based on using a single graph operator $\Zm$. 

The possibility of using \emph{multiple operators} was first observed in \cite{gavili2017shift}. Multiple graph operators $\Zc=\{\Zm^{(1)},\Zm^{(2)},\dots,\Zm^{(m)}\}$ that are jointly diagonalizable (i.e., have a common eigenbasis) can be obtained for both cycle graphs \cite{gavili2017shift} and line graphs \cite{lu2018efficient}. Essentially, those operators are by themselves graph filter matrices with different frequency responses. Thus, unlike \eqref{eq:pgf}, which is a polynomial of a single operator, we can design graph filters of the form:
\begin{equation}
\label{eq:mpgf}
\Hm_{\Zc,K}=p_K(\Zm^{(1)},\Zm^{(2)},\dots,\Zm^{(m)}),
\end{equation}
where $p_K(\cdot)$ stands for a multivariate polynomial with degree $K$ and arbitrary coefficients. 
Given the graph filter expression \eqref{eq:mpgf}, iterative algorithms for filter implementation have been recently studied in \cite{emirov2020polynomial}. Since $\Hm_{\{\Zm\},K}=p_K(\Zm)$ reduces to \eqref{eq:pgf}, the form \eqref{eq:mpgf} is a generalization of the PGF expression. We refer to \eqref{eq:mpgf} as \emph{multivariate polynomial graph filter (MPGF)}. 

In this paper, we focus on filtering operations based on the well-known discrete cosine transform (DCT)  and discrete sine transform (DST) \cite{strang1999discrete}, as well as their extension to all discrete trigonometric transforms (DTTs), i.e., 8 types of DCTs and 8 types of DSTs \cite{puschel2003algebraic}. All DTTs are GFTs associated with  uniform line graphs \cite{strang1999discrete,puschel2003algebraic}. DTT filters are based on the 
following operations: 1) computing the DTT of the input signal, 2) scaling each of the computed DTT coefficients, and 3) performing the inverse DTT. In particular, DCT filters \cite{chen1976image} have long been studied and are typically implemented using forward and inverse DCT.  As an alternative, we propose graph-based approaches to design and implement DTT  filters. The main advantage of graph based approaches is that they do not require  the DTT and inverse DTT steps, and instead can be applied directly in the signal domain, using suitable graph operators.  This allows us to define graph filtering approaches for all DTT filters, with applications including image resizing \cite{park2004design}, biomedical signal processing \cite{shin2010ideal}, medical imaging \cite{tuna2010gap}, and video coding \cite{zhang2015graph}.

Our work studies the design of efficient sample domain (graph vertex domain) graph filters, with particular focus on DTT filters. Specifically, for GFTs corresponding to any of the 16 DTTs, we derive a family  $\Zc$ of sparse graph operators with closed form expressions, which can be used in addition to the graph operator obtained from the well-known line graph model \cite{puschel2003algebraic}.
In this way, efficient DTT filters can be obtained using PGF and MPGF design approaches, yielding a lower complexity than a DTT filter implementation in the transform domain. Our main contributions are summarized as follows:
\begin{enumerate}
    \item We introduce multiple sparse graph operators specific to DTTs and allowing fast MPGF implementations. These sparse operators are DTT filters, which are special cases of graphs filters, but have not been considered in the general graph filtering  literature  \cite{shuman2018chebyshev,segarra2017optimal,loukas2015distributed,isufi2017autoregressive,liu2019filter,coutino2019advances}. While \cite{strang1999discrete} and \cite{puschel2003algebraic} establish the connection between DTTs and line graphs,  our proposed sparse graph operators for DTTs, which are no longer restricted to be line graphs, had not been studied in the literature.
    \item We introduce novel DTT filter design methods for graph vertex domain implementation. While in related work  \cite{chitprasert1990discrete,martucci1994symmetric,puschel2003algebraic}, DTT filtering  is typically performed  in the transform domain using convolution-multiplication properties, we introduce sample domain DTT filter implementations based on PGF and MPGF designs, and show that our designs with low degree polynomials lead to faster implementations as compared to those designs that require forward and inverse DTTs, especially in cases where DTT size is large. 
    \item In addition to the well-known least squares graph filter design, we propose a novel minimax design approach for both PGFs and MPGFs, which optimally minimizes the approximation error in terms of maximum absolute error in the graph frequencies.
    \item We provide novel insights on MPGF designs by demonstrating that using multiple operators leads to more efficient implementations, as compared to conventional PGF designs, for DTT filters with frequency responses that are non-smooth (e.g., ideal low-pass filters) or non-monotonic (e.g., bandpass filters). 
    \item We demonstrate experimentally the benefits of sparse DTT operators in image and video compression applications. In addition to filter operation, our approach can also be used to evaluate the transform domain weighted energy given by the Laplacian quadratic form, which has been used for rate-distortion optimization in the context of image and video coding \cite{hu2015multiresolution,fracastoro2020graph}. Following our recent work \cite{lu2018efficient}, we implement the proposed method in AV1, a real-world codec, where our method provides a speedup in the transform type search procedure. 
\end{enumerate} 

We highlight that, while \cite{emirov2020polynomial} studies MPGFs with a focus on distributed filter implementations, it does not investigate design approaches of MPGFs or how sparse operators for generic graphs can be obtained other than cycle and Cartesian product graphs. Our work complements the study in \cite{emirov2020polynomial} by considering 1) the case where GFT is a DTT, which corresponds to various line graphs, and 2) techniques to design MPGFs. In addition, the work presented in this paper is a more general framework than our prior work in \cite{lu2018efficient}, since the Laplacian quadratic form operation used in \cite{lu2018efficient} can be viewed as a special case of graph filtering operation. Furthermore, while our work in \cite{lu2018efficient} was restricted to DCT/ADST, in this paper we have extended these ideas to all DTTs. 

The rest of this paper is organized as follows. We review graph filtering concepts and some relevant properties of DTTs in Section~\ref{sec:preliminaries}. In Section~\ref{sec:dct_dst_operators}, we consider sparse operators for DTTs that can be obtained by extending well-known properties of DTTs. We also extend the results to 2D DTTs and provide some remarks on sparse operators for general graphs. Section~\ref{sec:sparse_operators} introduces PGF and MPGF design approaches using least squares and minimax criteria. An efficient filter design for Laplacian quadratic form approximation is also presented. 
Experimental results are shown in Section \ref{sec:experiment} to demonstrate the effectiveness of our methods in graph filter design as well as applications in video coding. Conclusions are given in Section~\ref{sec:conclusion}.

\section{Preliminaries}
\label{sec:preliminaries}

We start by reviewing relevant concepts in graph signal processing and DTTs. In what follows, entries in a matrix that are not displayed are meant to be zero.  Thus, the  order-reversal permutation matrix is:
\[
\Jm=\begin{pmatrix}
  &&& 1 \\ && 1 & \\ & \udots && \\ 1 &&&
\end{pmatrix},
\]
which satisfies $\Jm^\top=\Jm$ and $\Jm\Jm=\Id$, where the transpose of matrix $\Am$ is denoted as $\Am^\top$. The pseudo-inverse of $\Am$ is written as $\Am^\dagger$.

\subsection{Graph Fourier Transforms}
\label{subsec:gft}

Let $\Gc(\Vc,\Ec,\Wm)$ be an undirected graph with $N$ nodes and let $\xv$ be a length-$N$ graph signal associated to $\Gc$. Each node of $\Gc$ corresponds to an entry of $\xv$, and each edge $e_{ij}\in\Ec$ describes the inter-sample relation between nodes $i$ and $j$. The $(i,j)$ entry of the weight matrix, $w_{i,j}$, is the weight of the edge $e_{ij}$, and $\theta_i:=w_{i,i}$ is the weight of the self-loop on node $i$. Defining  $\Thetam=\text{diag}(\theta_1,\dots,\theta_N)$ and $\Dm=\text{diag}(d_1,\dots,d_N)$ as diagonal matrices of self-loop weights and node degrees, $d_i=\sum_{j=1}^N w_{i,j}$, respectively, the unnormalized and normalized graph Laplacian matrices are
\begin{equation}
\label{eq:def_lapl}
\Lm=\Dm-\Wm+\Thetam, \quad \Lcb=\Dm^{-1/2}\Lm\Dm^{-1/2}.
\end{equation}
In what follows, unless stated otherwise, we refer to the unnormalized version, $\Lm$, as the graph Laplacian. All graphs we consider are assumed to be undirected.

The graph Fourier transform (GFT) is obtained from the eigen-decomposition of the graph Laplacian, $\Lm=\Phim\Lambdam\Phim^\top$, with eigenvalues $\lambda_1\leq\cdots\leq\lambda_N$ in ascending order. 
The vector of GFT coefficients for graph signal $\xv$ is $\hat{\xv}=\Phim^\top\xv$. We note that the variation of signal $\xv$ on the graph can be measured by the Laplacian quadratic form:
\begin{equation}
\label{eq:lqf}
  \xv^\top\Lm\xv=\sum_{(i,j)\in\Ec}w_{i,j}(x_i-x_j)^2+\sum_{k=1}^N \theta_k x_k^2.
\end{equation}
The columns of $\Phim$, $\phiv_1,\dots,\phiv_N$ form an orthogonal basis and each of them can be viewed as a graph signal with variation equal to the associated eigenvalues $\lambda_1,\dots,\lambda_N$, which are called \emph{graph frequencies}.

\subsection{Graph Filters}
\label{subsec:graph_filter}
We consider a 1-hop graph operator $\Zm$, which could be the adjacency matrix or one of the Laplacian matrices. 
For a given signal $\xv$,  $\yv=\Zm\xv$ defines an operation where the output at each node is a function of values at its 1-hop neighbors (e.g., when $\Zm=\Am$, $y(i)=\sum_{j\in\Nc(i)} x(j)$, where $\Nc(i)$ is the set of nodes that are neighbors of $i$). Furthermore, it can be shown that $\yv=\Zm^K\xv$ is a $K$-hop operation, and thus for a degree-$K$ polynomial of $\Zm$, as in \eqref{eq:pgf}, the output at node $i$ depends on its $K$-hop neighbors. 
The operation in \eqref{eq:pgf} is thus called a \emph{graph filter}, an FIR graph filter, or a polynomial graph filter (PGF). In what follows, we refer to $\Zm$ as \emph{graph operator} for short\footnote{In the literature, $\Zm$ is often called \emph{graph shift operator} \cite{sandryhaila2013discrete,segarra2017optimal,gavili2017shift,isufi2017autoregressive}. Here, we simply call it graph operator, since its properties are different from shift in conventional signal processing, which is always reversible, while the graph operator $\Zm$, in most cases, is not.}. For the rest of this paper, we choose $\Zm=\Lm$ or define $\Zm$ as a matrix with the same eigenbasis as $\Lm$, e.g., $\Zm$ could be a polynomial of $\Lm$ such as $\Zm=2\Id-\Lm$. 

The matrix $\Phim$ of eigenvectors of $\Zm=\Lm$ is also the eigenvector matrix of any polynomial $\Hm$ in the form of \eqref{eq:pgf}. The eigenvalue $h(\lambda_j)$ of $\Hm$ associated to $\phiv_j$ is called the \emph{frequency response} of $\lambda_j$. Note that with $\yv=\Hm\xv$, in the GFT domain we have $\hat{\yv}=h(\Lambdam)\hat{\xv}$, meaning that the filter operator scales the signal component with $\lambda_j$ frequency by $h(\lambda_j)$ in the GFT domain. We also note that  \eqref{eq:graph_filter_freq} generalizes the notion of digital filter: when $\Phim$ is the discrete Fourier transform (DFT) matrix, $\Hm$ reduces to the classical Fourier filter \cite{sandryhaila2014discrete}. Given a desired graph frequency response $\hv=(h_1,\dots,h_N)^\top$, its associated polynomial coefficients in \eqref{eq:pgf} can be obtained by solving a least squares minimization problem \cite{sandryhaila2014discrete}:
\begin{equation}
\label{eq:fir_approx_problem}
\gv=\underset{\gv}{\text{argmin}} \quad || \hv - \Psim\gv ||^2, \quad\text{where }\Psim=
\begin{pmatrix}
1 & \lambda_1 & \dots & \lambda_1^K \\ 
\vdots & \vdots & \vdots & \vdots \\ 
1 & \lambda_N & \dots & \lambda_N^K
\end{pmatrix},
\end{equation}
with $\lambda_j$ being the $j$-th eigenvalue of $\Zm$. The PGF operation $\yv=\Hm\xv$ can be implemented efficiently by computing: 1) $\tv^{(0)}=g_K\xv$, 2) $\tv^{(i)}=\Zm\tv^{(i-1)}+g_{K-i}\xv$, and 3) $\yv=\tv^{(K)}$.
This algorithm does not require GFT computation, and its complexity depends on the degree $K$ and how sparse $\Zm$ is (with lower complexity for sparser $\Zm$).

\subsection{Discrete Cosine and Sine Transforms}
\label{subsec:dct_dst}

\begin{figure}
    \centering
    \subfigure[]{
    \includegraphics[width=.23\textwidth]{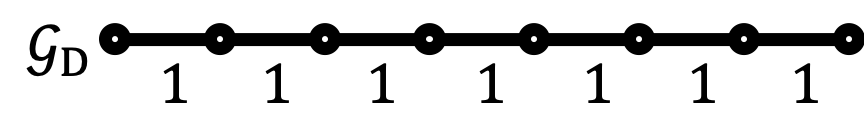}} %
    \subfigure[]{
    \includegraphics[width=.23\textwidth]{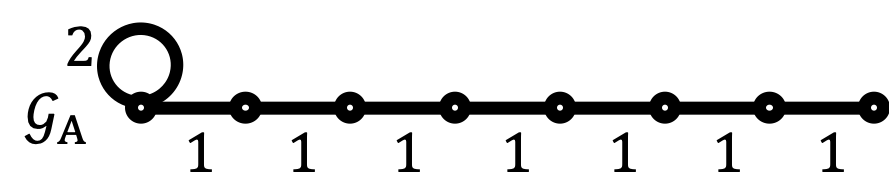}}
    \caption{Graphs associated to (a) DCT-II, (b) DST-IV (ADST).}
    \label{fig:dct_dst_graphs}
\end{figure}

The discrete cosine transform (DCT) and discrete sine transform (DST) are orthogonal transforms that operate on a finite vector, with basis functions derived from cosines and sines, respectively. Discrete trigonometric transforms (DTTs) comprise eight types of DCT and eight types of DST, which are defined depending on how samples are taken from continuous cosine and sine functions \cite{wang1985discrete,puschel2003algebraic}. We denote them by DCT-I to DCT-VIII, and DST-I to DST-VIII, and list their forms in Table~\ref{tab:eigenpairs}.

DCT and ADST are widely used in image and video coding. In this paper, we refer the terms ``DCT'' and ``ADST'' to DCT-II and DST-IV\footnote{DST-VII was shown to optimally decorrelate intra residual pixels under a Gaussian Markov model \cite{han2012jointly,hu2015intra-prediction}, but its variant DST-IV is amenable to fast implementations while experimentally achieving a similar coding efficiency  \cite{han2013butterfly}. In this paper, we refer to DST-IV as ADST, as in the AV1 codec \cite{chen2020overview}.}, respectively, unless stated otherwise. For $j=1,\dots,N$ and $k=1,\dots,N$, we denote the $k$-th element of the $j$-th length-$N$ DCT and ADST functions as
\begin{align}
\label{eq:dct}
  & \text{DCT-II:} \quad u_j(k)=\sqrt{\frac{2}{N}}c_j\cos \frac{(j-1)(k-\frac{1}{2})\pi}{N}, \\
\label{eq:dst4}
  & \text{DST-IV:} \quad v_j(k)=\sqrt{\frac{2}{N}}\sin \frac{(j-\frac{1}{2})(k-\frac{1}{2})\pi}{N}.
\end{align}
with normalization constant $c_j$ being $1/\sqrt{2}$ for $j=1$ and $1$ otherwise. 
If those basis functions are written in vector form $\uv_j,\vv_j\in\mathbb{R}^N$, it was pointed out in \cite{strang1999discrete} that the $\uv_j$ are eigenvectors of the Laplacian matrix $\Lm_\text{D}$, and in \cite{puschel2003algebraic} that $\vv_j$ are eigenvectors of $\Lm_\text{A}$, with
\begin{equation}
\label{eq:Ld}
  \scriptsize
  \Lm_\text{D}=\begin{pmatrix}
    1 & -1 & & & \\
    -1 & 2 & -1 & & \\
    & \ddots & \ddots & \ddots & \\
    & & -1 & 2 & -1 \\
    & & & -1 & 1 
  \end{pmatrix}, \quad \scriptsize\Lm_\text{A}=\begin{pmatrix}
    3 & -1 & & & \\
    -1 & 2 & -1 & & \\
    & \ddots & \ddots & \ddots & \\
    & & -1 & 2 & -1 \\
    & & & -1 & 1 
  \end{pmatrix}. 
\end{equation}
This means that the DCT and ADST are GFTs corresponding to Laplacian matrices $\Lm_\text{D}$ and $\Lm_\text{A}$, respectively. Their associated graphs $\Gc_\text{D}$ and $\Gc_\text{A}$ with $N=6$ are shown in Figs.~\ref{fig:dct_dst_graphs}(a) and (b). The eigenvalues of $\Lm_\text{D}$ corresponding to $\uv_j$ are $\omega_j=2-2\cos ((j-1)\pi/N)$, and those of $\Lm_\text{A}$ corresponding to $\vv_j$ are $\delta_j=2-2\cos((j-1/2)\pi/N)$.

\section{Sparse DCT and DST Operators}
\label{sec:dct_dst_operators}

Classical PGFs can be extended to MPGFs \cite{emirov2020polynomial} if multiple graph operators are available \cite{gavili2017shift}. Let $\Lm=\Phim\Lambdam\Phim^\top$ be a Laplacian with GFT $\Phim$ and assume  we have a series of graph operators $\Zc=\{\Zm^{(k)}\}_{k=1}^M$ that share the same eigenvectors as $\Lm$, but with different eigenvalues:
\begin{align*}
  \Zm^{(k)}=\Phim \Lambdam^{(k)} \Phim^\top, \quad
  \Lambdam^{(k)}=\diag(\lambdav^{(k)})=\diag(\lambda_1^{(k)},\dots,\lambda_N^{(k)}),
\end{align*}
where $\lambdav^{(k)}=(\lambda_1^{(k)},\dots,\lambda_N^{(k)})^\top$ denotes the vector of eigenvalues of $\Zm^{(k)}$. When the polynomial degree is $K=1$ in \eqref{eq:mpgf}, we have:
\begin{equation}
\label{eq:mpgf_k1}
  \Hm_{\Zc,1} = g_0\Id + \sum_{m=1}^M g_m \Zm^{(m)},
\end{equation}
where $g_k$ are coefficients. When $K=2$, we have
\begin{align}
\label{eq:mpgf_k2}
  \Hm_{\Zc,2} &= g_0\Id + \sum_{m=1}^M g_m \Zm^{(m)} \nonumber\\
  & \quad + g_{M+1}\Zm^{(1)}\Zm^{(1)} + g_{M+2}\Zm^{(1)}\Zm^{(2)} + \dots + g_{2M}\Zm^{(1)}\Zm^{(M)} \nonumber\\
  & \quad + g_{2M+1} \Zm^{(2)}\Zm^{(2)} + \dots + g_{3M-1} \Zm^{(2)}\Zm^{(M)} \nonumber\\
  & \quad + \dots \nonumber\\
  & \quad + g_{(M^2+3M)/2}\Zm^{(M)}\Zm^{(M)},
\end{align}
where the terms $\Zm^{(j)}\Zm^{(i)}$ with $j>i$ are not required in \eqref{eq:mpgf_k2} because all operators commute, i.e., $\Zm^{(i)}\Zm^{(j)}=\Zm^{(j)}\Zm^{(i)}$. Expressions with a higher degree can be obtained with polynomial kernel expansion \cite{hofmann2008kernel}. We also note that, since $\Hm_{\{\Zm\},K}$ reduces to the form of $\Hm$ in \eqref{eq:pgf}, $\Hm_{\Zc,K}$ is a generalization of PGF and thus provides more degrees of freedom for the filter design procedure.

As pointed out in the introduction, DTT filters are essentially graph filters. This means that they can be implemented with PGFs as in \eqref{eq:pgf}, without applying any forward or inverse DTT. Next, we will go one step further by introducing multiple sparse operators for each DTT, which allows the implementation of DTT filters using MPGFs.

\renewcommand{\arraystretch}{2}
\begin{table}
\centering
\caption{Definitions of DTTs and the eigenvalues of their sparse operators. The indices $j$ and $k$ range from $1$ to $N$. Scaling factors for rows and columns are given by $c_j=1/\sqrt{2}$ for $j=1$ and $1$ otherwise, and $d_j=1/\sqrt{2}$ for $j=N$ and $1$ otherwise.}
\label{tab:eigenpairs}
\begin{tabular}{lll}
\hline
\multirow{2}{*}{DTT} & \multirow{2}{*}{Transform functions $\phi_j(k)$} & Eigenvalue of $\Zm^{(\ell)}$\\ & & associated to $\phiv_j$ \\
\hline
DCT-I & $\sqrt{\frac{2}{N-1}}c_j c_k d_j d_k \cos\frac{(j-1)(k-1)\pi}{N-1}$ & 
$2\cos\left(\frac{\ell(j-1)\pi}{N-1}\right)$ \\
DCT-II & $\sqrt{\frac{2}{N}} c_j\cos \frac{(j-1)(k-1/2)\pi}{N}$ & 
$2\cos\left(\frac{\ell(j-1)\pi}{N}\right)$ \\
DCT-III & $\sqrt{\frac{2}{N}}c_k\cos \frac{(j-1/2)(k-1)\pi}{N}$ & 
$2\cos\left(\frac{\ell(j-1/2)\pi}{N}\right)$ \\
DCT-IV & $\sqrt{\frac{2}{N}} \cos \frac{(j-1/2)(k-1/2)\pi}{N}$ & 
$2\cos\left(\frac{\ell(j-1/2)\pi}{N}\right)$ \\
DCT-V & $\frac{2}{\sqrt{2N-1}} c_j c_k\cos \frac{(j-1)(k-1)\pi}{N-1/2}$ & 
$2\cos\left(\frac{\ell(j-1)\pi}{N-1/2}\right)$ \\
DCT-VI & $\frac{2}{\sqrt{2N-1}} c_j d_k\cos \frac{(j-1)(k-1/2)\pi}{N-1/2}$ & 
$2\cos\left(\frac{\ell(j-1)\pi}{N-1/2}\right)$ \\
DCT-VII & $\frac{2}{\sqrt{2N-1}} d_j c_k\cos \frac{(j-1/2)(k-1)\pi}{N-1/2}$ & 
$2\cos\left(\frac{\ell(j-1/2)\pi}{N-1/2}\right)$ \\
DCT-VIII & $\frac{2}{\sqrt{2N+1}}\cos \frac{(j-1/2)(k-1/2)\pi}{N+1/2}$ & 
$2\cos\left(\frac{\ell(j-1/2)\pi}{N+1/2}\right)$ \\
DST-I & $\sqrt{\frac{2}{N+1}}\sin \frac{jk\pi}{N+1}$ & 
$2\cos\left(\frac{\ell j\pi}{N+1}\right)$ \\
DST-II & $\sqrt{\frac{2}{N}} d_j \sin \frac{j(k-1/2)\pi}{N}$ & 
$2\cos\left(\frac{\ell j\pi}{N}\right)$ \\
DST-III & $\sqrt{\frac{2}{N}} d_k\sin \frac{(j-1/2)k\pi}{N}$ & 
$2\cos\left(\frac{\ell (j-1/2)\pi}{N}\right)$ \\
DST-IV & $\sqrt{\frac{2}{N}}\sin \frac{(j-1/2)(k-1/2)\pi}{N}$ & 
$2\cos\left(\frac{\ell(j-1/2)\pi}{N}\right)$ \\
DST-V & $\frac{2}{\sqrt{2N+1}}\sin \frac{jk\pi}{N+1/2}$ & 
$2\cos\left(\frac{\ell j\pi}{N+1/2}\right)$ \\
DST-VI & $\frac{2}{\sqrt{2N+1}}\sin \frac{j(k-1/2)\pi}{N+1/2}$ & 
$2\cos\left(\frac{\ell j\pi}{N+1/2}\right)$ \\
DST-VII & $\frac{2}{\sqrt{2N+1}}\sin \frac{(j-1/2)k\pi}{N+1/2}$ & 
$2\cos\left(\frac{\ell(j-1/2)\pi}{N+1/2}\right)$ \\
DST-VIII & $\frac{2}{\sqrt{2N-1}} d_j d_k\sin \frac{(j-1/2)(k-1/2)\pi}{N-1/2}$ & 
$2\cos\left(\frac{\ell(j-1/2)\pi}{N-1/2}\right)$ \\
\hline
\end{tabular}
\end{table}
\renewcommand{\arraystretch}{1}

The use of polynomial \eqref{eq:pgf} to perform filtering in the vertex domain, rather in the frequency domain, is advantageous only if the operator is sparse. In this section, our main goal is to show that multiple \textit{sparse} operators can be found for DTTs. First, the result of \cite{strang1999discrete} will be generalized in Sec.~\ref{subsec:sparse_dct_filters} to derive multiple sparse operators from a single operator for DCT-II. A toy example for those operators is provided in Sec.~\ref{subsec:example_length4_dct2}. Then, in Sec.~\ref{subsec:filter_sparse_all}, we further show that, in addition to DCT-II, operators can be derived for all 16 DTTs based on the approach in Sec.~\ref{subsec:sparse_dct_filters}. Finally, sparse operators associated to 2D DTTs are presented in Sec.~\ref{subsec:sparse_2d_filters}.

\subsection{Sparse DCT-II Operators}
\label{subsec:sparse_dct_filters}

Let $\uv_j$ denote the DCT basis vector with entries from \eqref{eq:dct} and let  $\Lm_\text{D}$ be the Laplacian of a uniform line graph, \eqref{eq:Ld}. The following proposition from \cite{strang1999discrete} and its proof, developed for the line graph case, will be useful to find additional sparse operators:
\begin{proposition}[\cite{strang1999discrete}]
\label{prop:dct2}
$\uv_j$ is an eigenvector of $\Lm_\text{D}$ with eigenvalue $\omega_j=2-2\cos((j-1)\pi/N)$ for each $j=1,\dots N$
\end{proposition}
\noindent{\it Proof:}
It suffices to show an equivalent equation: $\Zm_\text{DCT-II}\cdot\uv_j=(2-\omega_j)\uv_j$, where 
\begin{equation}
\label{eq:Bd1}
  \Zm_\text{DCT-II}=2\Id-\Lm_\text{D}=\begin{pmatrix}
    1 & 1 & & & \\
    1 & 0 & 1 & & \\
    & \ddots & \ddots & \ddots & \\
    & & 1 & 0 & 1 \\
    & & & 1 & 1 
  \end{pmatrix}.
\end{equation}
For $1\leq p \leq N$, the $p$-th element of $\Zm_\text{DCT-II}\cdot\uv_j$ is
\[
  (\Zm_\text{DCT-II}\cdot\uv_j)_p = \left\{ 
  \begin{array}{ll}
  u_j(1)+u_j(2), & p=1 \\
  u_j(p-1)+u_j(p+1), & 2\leq p \leq N-1 \\
  u_j(N-1)+u_j(N), & p=N.
  \end{array}\right.
\]
Following the expression in \eqref{eq:dct}, we extend the definition of $u_j(k)$ to an arbitrary integer $k$. The even symmetry of the cosine function at $0$ and $\pi$ gives $u_j(0)=u_j(1)$ and $u_j(N)=u_j(N+1)$, and thus
\begin{align}
\label{eq:indices_symmetry}
  (\Zm_\text{DCT-II}\cdot\uv_j)_1 &= u_j(0)+u_j(2), \nonumber\\
  (\Zm_\text{DCT-II}\cdot\uv_j)_N &= u_j(N-1)+u_j(N+1).
\end{align}
This means that for all $p=1,\dots,N$, 
\begin{subequations}
\begin{align}
\label{eq:recursive-1}
  & (\Zm_\text{DCT-II}\cdot\uv_j)_p =u_j(p-1)+u_j(p+1) \\
\label{eq:recursive-2}
  &=\sqrt{\frac{2}{N}} c_j \left[\cos \frac{(j-1)(p-\frac{3}{2})\pi}{N}+\cos \frac{(j-1)(p+\frac{1}{2})\pi}{N} \right]\\
\label{eq:recursive-3}
  &=2\sqrt{\frac{2}{N}} c_j\cos\frac{(j-1)(p-\frac{1}{2})\pi}{N}\cos\frac{(j-1)\pi}{N} \\
\label{eq:recursive-4}
  &= (2-\omega_j) u_j(p),
\end{align}
\end{subequations}
which verifies $\Zm_\text{DCT-II}\cdot\uv_j=(2-\omega_j)\uv_j$. 
Note that in \eqref{eq:recursive-2}, we have applied the sum-to-product trigonometric identity: 
\begin{equation}
\label{eq:sum2prod_cosine}
\pushQED{\qed}
  \cos\alpha+\cos\beta=2\cos\left(\frac{\alpha+\beta}{2}\right)
  \cos\left(\frac{\alpha-\beta}{2}\right).
\popQED
\end{equation}

Now we can extend the above result as follows. When $u_j(q\pm 1)$ is replaced by $u_j(q\pm \ell)$ in \eqref{eq:recursive-1}, this identity also applies, which generalizes \eqref{eq:recursive-1}-\eqref{eq:recursive-4} to
\begin{align}
\label{eq:recursive_2}
  & u_j(p-\ell)+u_j(p+\ell) \nonumber\\
  &=\sqrt{\frac{2}{N}} c_j\left[ \cos \frac{(j-1)(p-\ell-\frac{1}{2})\pi}{N}+\cos \frac{(j-1)(p+\ell-\frac{1}{2})\pi}{N} \right] \nonumber\\
  &=2\sqrt{\frac{2}{N}}c_j\cos\frac{(j-1)(p-\frac{1}{2})\pi}{N}\cos\frac{\ell(j-1)\pi}{N} \nonumber\\
  &= \left(2\cos\frac{\ell(j-1)\pi}{N}\right) u_j(p).
\end{align}
As in \eqref{eq:indices_symmetry}, we can apply even symmetry of the cosine function at 0 and $\pi$, to replace indices $p-\ell$ or $p+\ell$ that are out of the range $[1,N]$ by those within the range:
\begin{align*}
  & u_j(p-\ell) = u_j(-p+\ell+1), \\
  & u_j(p+\ell) = u_j(-p-\ell+2N+1).
\end{align*}

Then, an $N\times N$ matrix $\Zm_\text{DCT-II}^{(\ell)}$ can be defined such that the left hand side of \eqref{eq:recursive_2} corresponds to $(\Zm_\text{DCT-II}^{(\ell)}\cdot\uv_j)_p$. This leads to the following proposition:
\begin{proposition}
\label{prop:dct}
For $\ell=1,\dots,N-1$, we define $\Zm_\text{DCT-II}^{(\ell)}$ as a $N\times N$ matrix, whose $p$-th row has only two non-zero elements specified as follows:
\begin{align*}
  & {\left(\Zm_\text{DCT-II}^{(\ell)}\right)}_{p,{q_1}}=1, \quad q_1 = \left\{\begin{array}{ll}
  p-\ell, & \text{ if } p-\ell \geq 1\\
  -p+\ell+1, & \text{ otherwise}
  \end{array}\right. \\
  & {\left(\Zm_\text{DCT-II}^{(\ell)}\right)}_{p,{q_2}}=1, \quad q_2 = \left\{\begin{array}{ll}
  p+\ell, & \text{ if } p+\ell \leq N\\
  -p-\ell+2N+1, & \text{ otherwise}
  \end{array}\right. 
\end{align*}
This matrix $\Zm_\text{DCT-II}^{(\ell)}$ has eigenvectors $\uv_j$ with associated eigenvalues $2\cos(\ell(j-1)\pi/N)$ for $j=1,\dots,N$. 
\end{proposition}

\begin{figure}
    \centering
    \subfigure[$\Zm_{\text{DCT-II}}^{(1)}$, $\Zm_{\text{DCT-II}}^{(2)}$, $\Zm_{\text{DCT-II}}^{(3)}$, and $\Zm_{\text{DCT-II}}^{(4)}$]{
    $\scriptsize \begin{pmatrix}
    1&1&0&0 \\ 1&0&1&0 \\ 0&1&0&1 \\ 0&0&1&1
    \end{pmatrix}
    \scriptsize \begin{pmatrix}
    0&1&1&0 \\ 1&0&0&1 \\ 1&0&0&1 \\ 0&1&1&0
    \end{pmatrix}
    \scriptsize \begin{pmatrix}
    0&0&1&1 \\ 0&1&0&1 \\ 1&0&1&0 \\ 1&1&0&0
    \end{pmatrix}
    \scriptsize \begin{pmatrix}
    0&0&0&2 \\ 0&0&2&0 \\ 0&2&0&0 \\ 2&0&0&0
    \end{pmatrix}$}\\
    \setlength\arraycolsep{2pt}
    \subfigure[$\Lm_{\text{DCT-II}}^{(1)}$, $\Lm_{\text{DCT-II}}^{(2)}$, $\Lm_{\text{DCT-II}}^{(3)}$, and $\Lm_{\text{DCT-II}}^{(4)}$]{
    $\scriptsize \begin{pmatrix}
    1&-1&0&0 \\ -1&2&-1&0 \\ 0&-1&2&-1 \\ 0&0&-1&1
    \end{pmatrix}
    \scriptsize \begin{pmatrix}
    2&-1&-1&0 \\ -1&2&0&-1 \\ -1&0&2&-1 \\ 0&-1&-1&2
    \end{pmatrix}
    \scriptsize \begin{pmatrix}
    2&0&-1&-1 \\ 0&1&0&-1 \\ -1&0&1&0 \\ -1&-1&0&2
    \end{pmatrix}
    \scriptsize \begin{pmatrix}
    2&0&0&-2 \\ 0&2&-2&0 \\ 0&-2&2&0 \\ -2&0&0&2
    \end{pmatrix}$}
    \subfigure[$\Gc_{\text{DCT-II}}^{(1)}$, $\Gc_{\text{DCT-II}}^{(2)}$, $\Gc_{\text{DCT-II}}^{(3)}$, and $\Gc_{\text{DCT-II}}^{(4)}$]{
    \includegraphics[width=.48\textwidth]{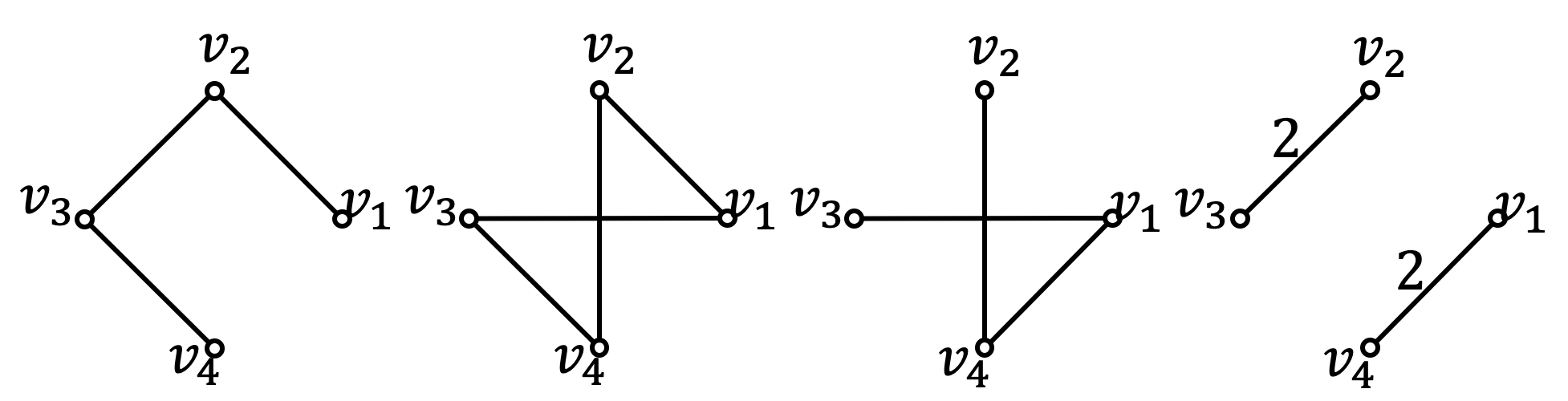}}
    \caption{(a) Sparse operators $\Zm_{\text{DCT-II}}^{(j)}$, (b) their associated Laplacian matrices $\Lm_{\text{DCT-II}}^{(j)}=2\Id-\Zm_{\text{DCT-II}}^{(j)}$, and (c) associated graphs $\Gc^{(j)}$ for the length-4 DCT-II.}
    \label{fig:dct2_operators_graphs}
\end{figure}

\begin{figure*}[t]
\begin{minipage}{.46\textwidth}
\subfigure[$\Zm_\text{DCT-I}^{(1)}$ to $\Zm_\text{DCT-I}^{(5)}$]{
\includegraphics[height=.055\textheight]{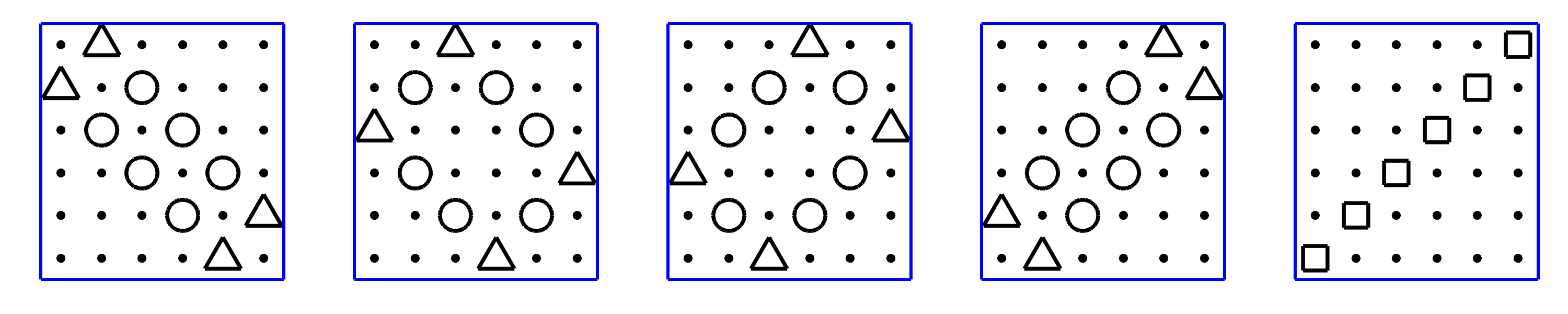}}
\end{minipage}\hfill
\begin{minipage}{.54\textwidth}
\subfigure[$\Zm_\text{DCT-II}^{(1)}$ to $\Zm_\text{DCT-II}^{(6)}$]{
\includegraphics[height=.055\textheight]{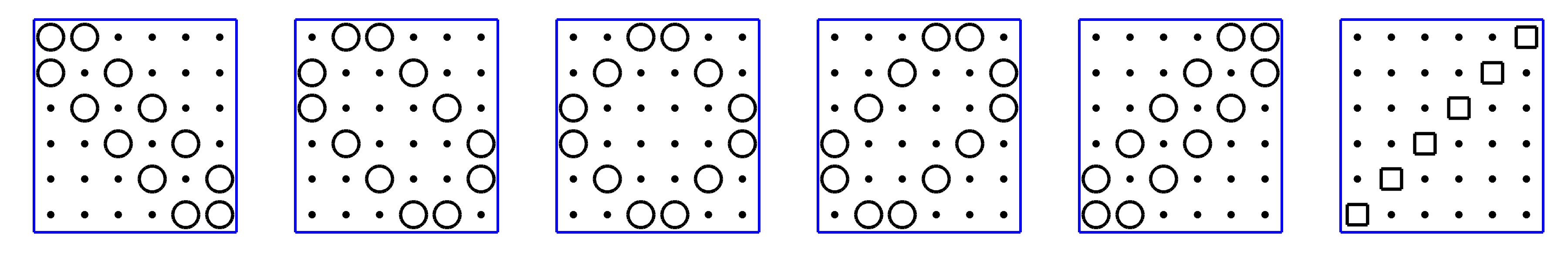}}
\end{minipage}
\begin{minipage}{.46\textwidth}
\subfigure[$\Zm_\text{DCT-III}^{(1)}$ to $\Zm_\text{DCT-III}^{(5)}$]{
\includegraphics[height=.055\textheight]{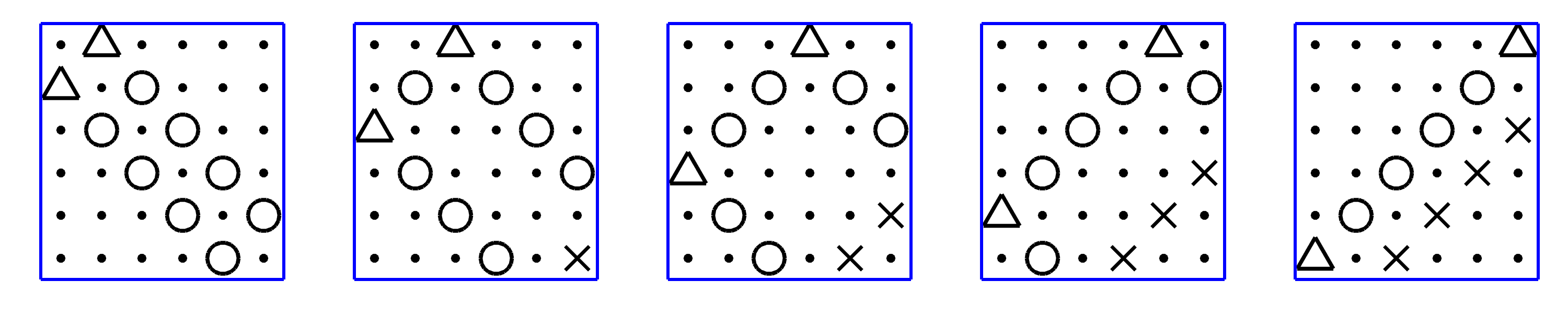}}
\end{minipage}\hfill
\begin{minipage}{.54\textwidth}
\subfigure[$\Zm_\text{DCT-IV}^{(1)}$ to $\Zm_\text{DCT-IV}^{(5)}$]{
\includegraphics[height=.055\textheight]{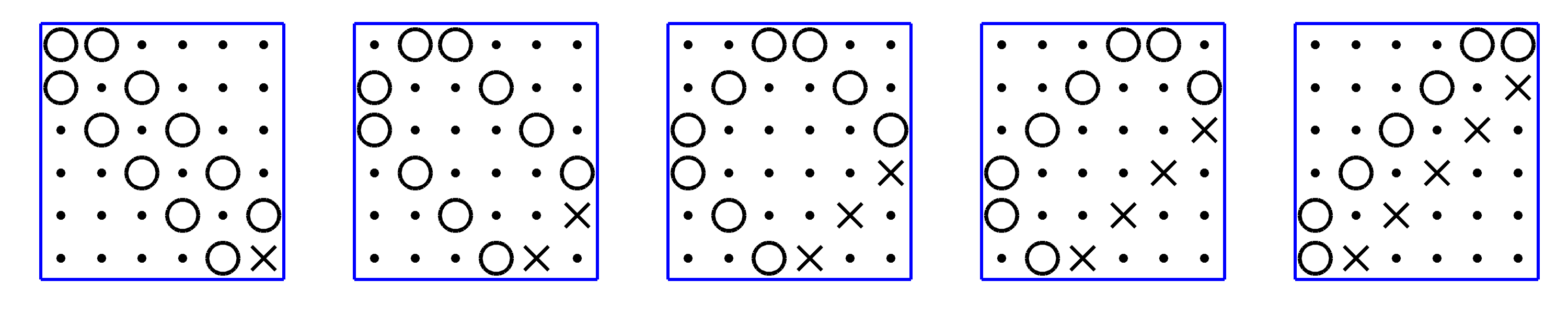}}
\end{minipage}
\begin{minipage}{.46\textwidth}
\subfigure[$\Zm_\text{DCT-V}^{(1)}$ to $\Zm_\text{DCT-V}^{(5)}$]{
\includegraphics[height=.055\textheight]{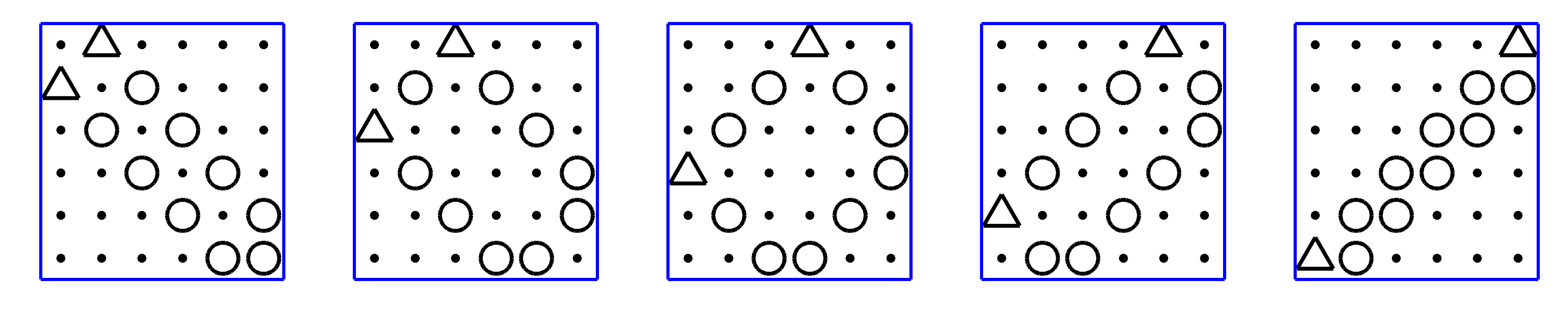}}
\end{minipage}\hfill
\begin{minipage}{.54\textwidth}
\subfigure[$\Zm_\text{DCT-VI}^{(1)}$ to $\Zm_\text{DCT-VI}^{(5)}$]{
\includegraphics[height=.055\textheight]{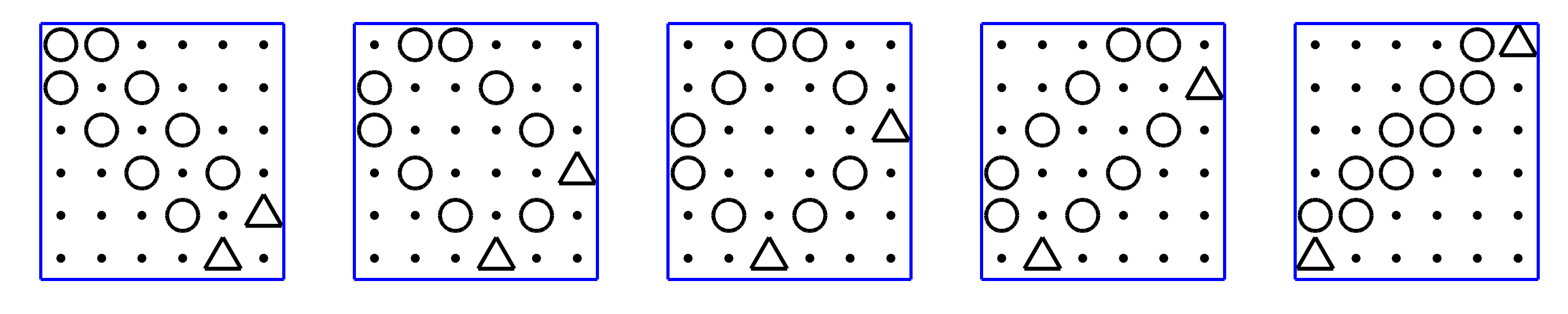}}
\end{minipage}
\begin{minipage}{.46\textwidth}
\subfigure[$\Zm_\text{DCT-VII}^{(1)}$ to $\Zm_\text{DCT-VII}^{(5)}$]{
\includegraphics[height=.055\textheight]{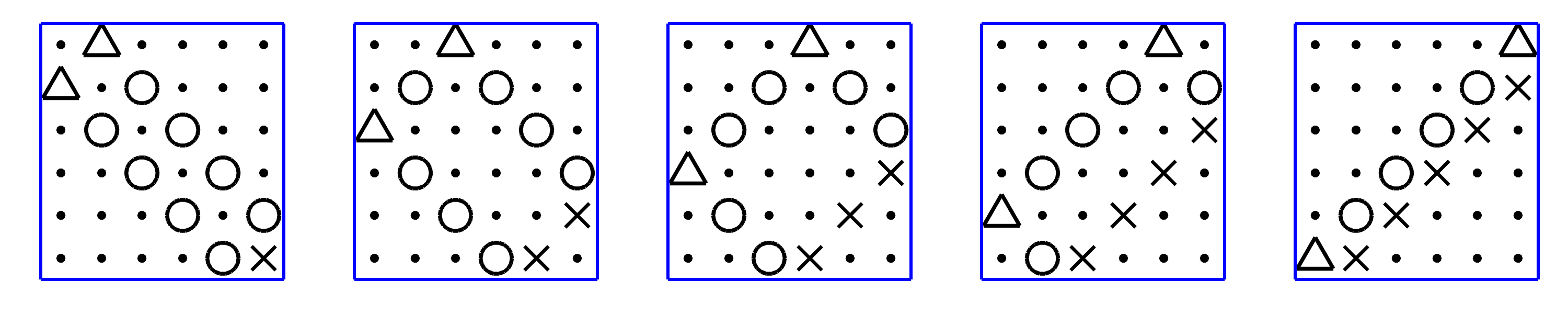}}
\end{minipage}\hfill
\begin{minipage}{.54\textwidth}
\subfigure[$\Zm_\text{DCT-VIII}^{(1)}$ to $\Zm_\text{DCT-VIII}^{(6)}$]{
\includegraphics[height=.055\textheight]{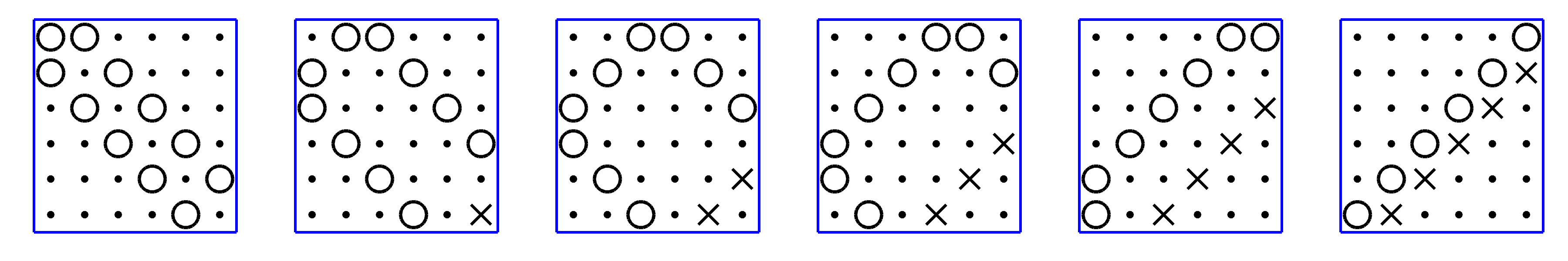}}
\end{minipage}
\caption{Sparse graph operators with length $N=6$ that associated to DCT-I to DCT-VIII. Different symbols represent different values: $\times=-1$, $\mathbf{\cdot}=0$, $\medcircle=1$, $\bigtriangleup=\sqrt{2}$, and $\square=2$.}
\label{fig:Ls_dct}
\end{figure*}

Note that $\Zm_\text{DCT-II}^{(1)}=\Zm_\text{DCT-II}$ as in \eqref{eq:Bd1}. Taking $\ell=2$ and $\ell=3$ and following Proposition~\ref{prop:dct}, we see that nonzero elements in $\Zm_\text{DCT-II}^{(2)}$ and $\Zm_\text{DCT-II}^{(3)}$ form rectangle-like patterns similar to that in $\Zm_\text{DCT-II}$:
\begin{equation}
\label{eq:z2_z3_dct2}
  \scriptsize
  \Zm_\text{DCT-II}^{(2)}=\begin{pmatrix}
    & 1 & 1 & & & \\
    1 & & & 1 & & \\
    1 & & & & \ddots & \\
    & 1 & & & & 1 \\
    & & \ddots & & & 1 \\
    & & & 1 & 1 & \\
  \end{pmatrix}, \hspace{.1cm} 
  \Zm_\text{DCT-II}^{(3)}=\begin{pmatrix}
    & & 1 & 1 & & \\
    & 1 & & & \ddots & \\
    1 & & & & & 1 \\
    1 & & & & & 1 \\
    & \ddots & & & 1 & \\
    & & 1 & 1 & & \\
  \end{pmatrix}
\end{equation}
For $\ell=N$, the derivations in \eqref{eq:recursive_2} are also valid, but with $\Zm_\text{DCT-II}^{(N)}=2\Jm$. The rectangular patterns we observe in \eqref{eq:z2_z3_dct2} can be simply extended to any arbitrary transform length $N$ (e.g., all such operators with $N=6$ are shown in Fig.~\ref{fig:Ls_dct}(b)). We also show the associated eigenvalues of $\Zm_\text{DCT-II}^{(\ell)}$ with arbitrary $N$ in Table~\ref{tab:eigenpairs}. Note that all the operators and their associated graphs are sparse. In particular, each operator has at most $2N$ non-zero entries and its corresponding graph has at most $N-1$ edges.

\subsection{Example--Length 4 DCT-II Operators}
\label{subsec:example_length4_dct2}

We show in Fig.~\ref{fig:dct2_operators_graphs}(a) all sparse operators $\Zm_{\text{DCT-II}}^{(\ell)}$ of DCT-II for $N=4$. In fact, those matrices can be regarded as standard operators on different graphs: by defining $\Lm_{\text{DCT-II}}^{(\ell)}=2\Id-\Zm_{\text{DCT-II}}^{(\ell)}$, we can view $\Lm_{\text{DCT-II}}^{(\ell)}$ as a Laplacian matrix of a different graph $\Gc_{\text{DCT-II}}^{(\ell)}$. For example, all the resulting $\Lm_{\text{DCT-II}}^{(\ell)}$'s and $\Gc_{\text{DCT-II}}^{(\ell)}$'s for a length-4 DCT-II are shown in Figure~\ref{fig:dct2_operators_graphs}(b) and (c), respectively. The rectangular patterns we observe in \eqref{eq:z2_z3_dct2} can be simply extended to any arbitrary transform length $N$. We also show the associated eigenvalues of $\Zm_\text{DCT-II}^{(\ell)}$ with arbitrary $N$ in Table~\ref{tab:eigenpairs}. 

We observe that, among all graphs in Fig.~\ref{fig:dct2_operators_graphs}(c), $\Gc_\text{DCT-II}^{(4)}$ is a disconnected graph with two connected components. It is associated to the operator
\[
  \Zm_\text{DCT-II}^{(4)}=\Phim_\text{DCT-II}\cdot\diag(2,-2,2,-2)\cdot\Phim_\text{DCT-II}^\top.
\]
Note that, while $\Zm_\text{DCT-II}^{(4)}$ is associated to a disconnected graph, it can still be used as a graph operator for DCT-II filter because it is diagonalized by $\Phim_\text{DCT-II}$. However, $\Zm_\text{DCT-II}^{(4)}$, as well as its polynomials, have eigenvalues with multiplicity 2. This means that a filter whose frequency response has distinct values (e.g. low-pass filter with $h(\lambda_1)>\dots>h(\lambda_4)$ cannot be realized as a PGF of $\Zm_\text{DCT-II}^{(4)}$). 

Based on the previous observation, we can see that those operators associated to disconnected graphs, and those having eigenvalues with high multiplicities lead to fewer degrees of freedoms in PGF and MPGF filter designs, as compared to an operators with distinct eigenvalues such as $\Zm_\text{DCT-II}^{(1)}$.


\begin{figure*}[t]
\begin{minipage}{.53\textwidth}
\subfigure[$\Zm_\text{DST-I}^{(1)}$ to $\Zm_\text{DST-I}^{(7)}$]{
\includegraphics[height=.055\textheight]{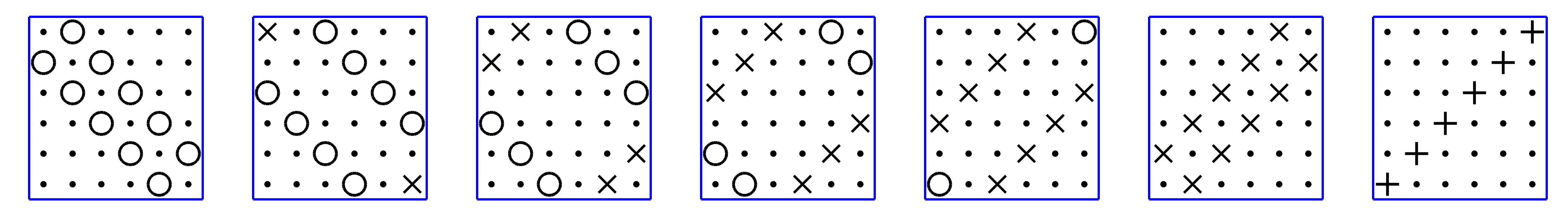}}
\end{minipage}\hfill
\begin{minipage}{.47\textwidth}
\subfigure[$\Zm_\text{DST-II}^{(1)}$ to $\Zm_\text{DST-II}^{(6)}$]{
\includegraphics[height=.055\textheight]{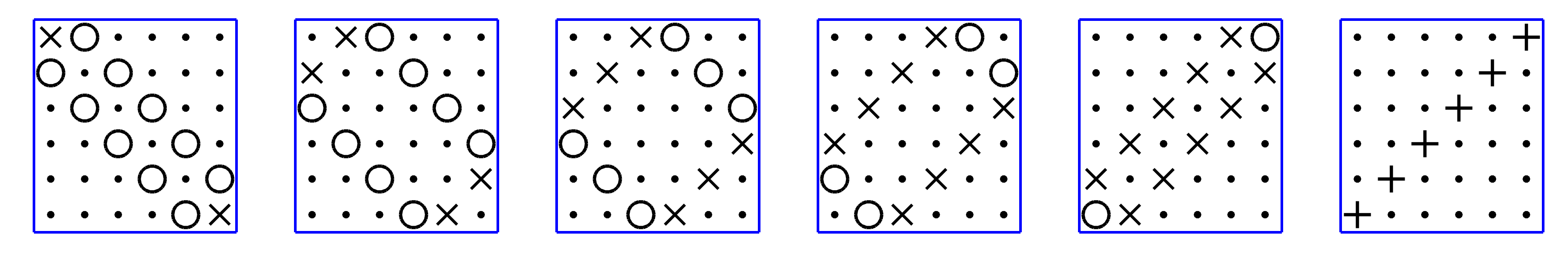}}
\end{minipage}
\begin{minipage}{.53\textwidth}
\subfigure[$\Zm_\text{DST-III}^{(1)}$ to $\Zm_\text{DST-III}^{(5)}$]{
\includegraphics[height=.055\textheight]{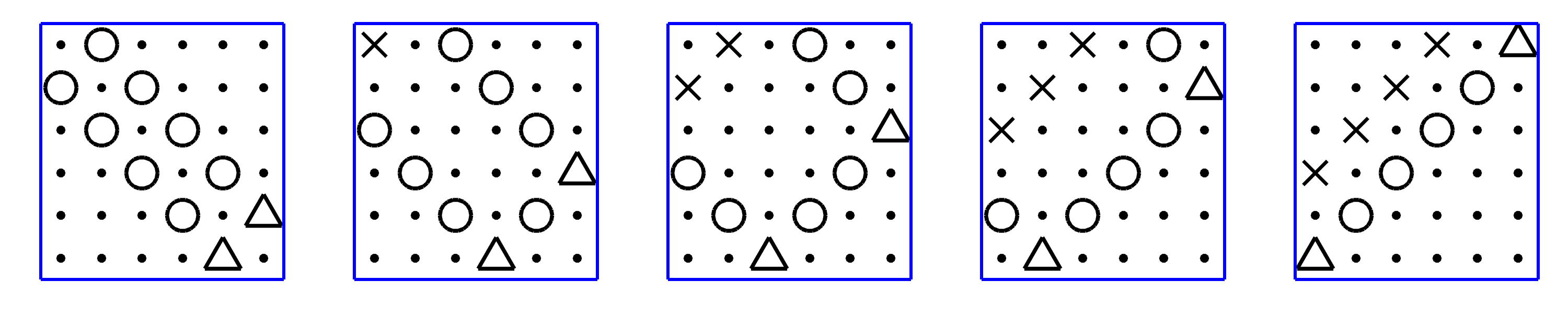}}
\end{minipage}\hfill
\begin{minipage}{.47\textwidth}
\subfigure[$\Zm_\text{DST-IV}^{(1)}$ to $\Zm_\text{DST-IV}^{(5)}$]{
\includegraphics[height=.055\textheight]{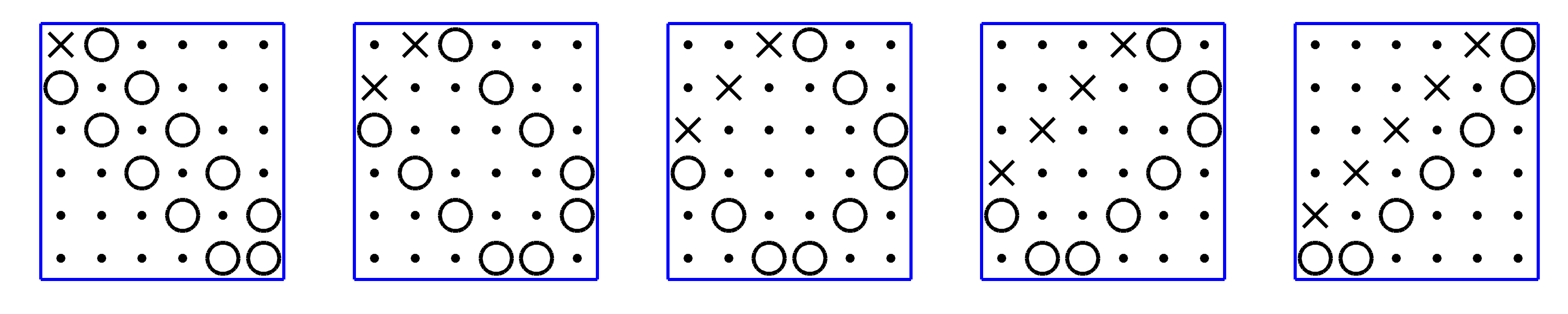}}
\end{minipage}
\begin{minipage}{.53\textwidth}
\subfigure[$\Zm_\text{DST-V}^{(1)}$ to $\Zm_\text{DST-V}^{(6)}$]{
\includegraphics[height=.055\textheight]{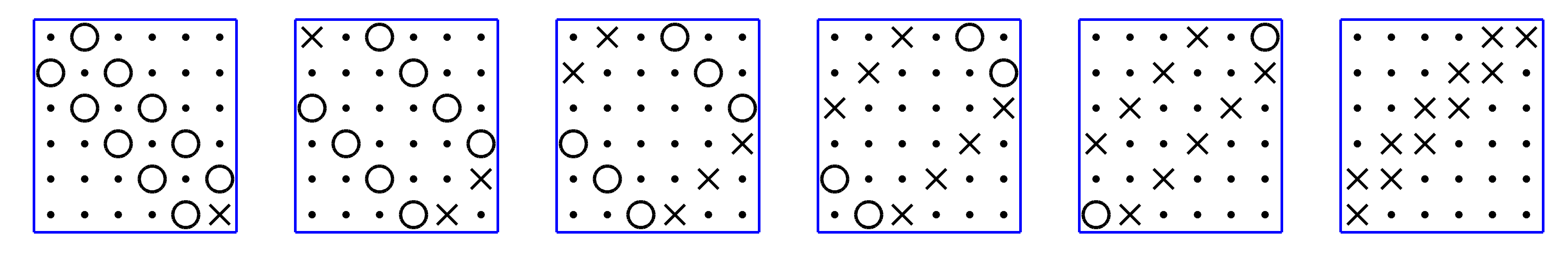}}
\end{minipage}\hfill
\begin{minipage}{.47\textwidth}
\subfigure[$\Zm_\text{DST-VI}^{(1)}$ to $\Zm_\text{DST-VI}^{(6)}$]{
\includegraphics[height=.055\textheight]{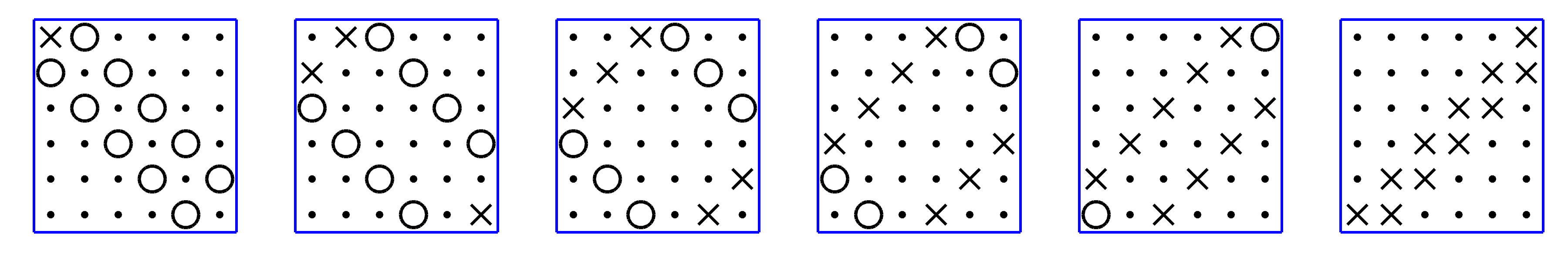}}
\end{minipage}
\begin{minipage}{.53\textwidth}
\subfigure[$\Zm_\text{DST-VII}^{(1)}$ to $\Zm_\text{DST-VII}^{(6)}$]{
\includegraphics[height=.055\textheight]{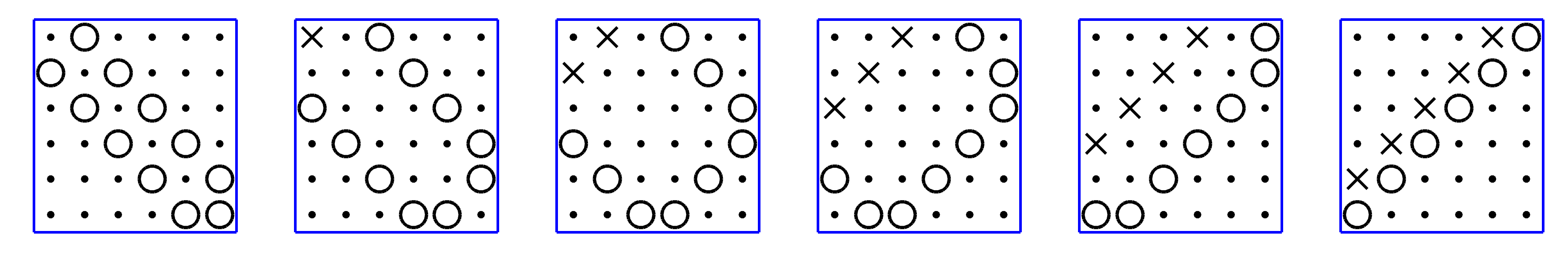}}
\end{minipage}\hfill
\begin{minipage}{.47\textwidth}
\subfigure[$\Zm_\text{DST-VIII}^{(1)}$ to $\Zm_\text{DST-VIII}^{(5)}$]{
\includegraphics[height=.055\textheight]{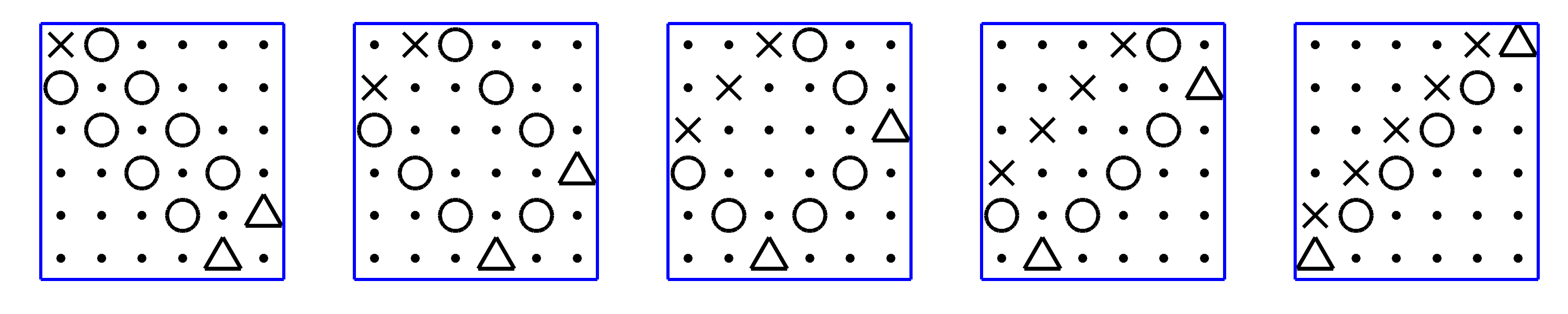}}
\end{minipage}
\caption{Sparse graph operators with length $N=6$ that associated to DST-I to DST-VIII. Different symbols represent different values: $+=-2$, $\times=-1$, $\mathbf{\cdot}=0$, $\medcircle=1$, and $\bigtriangleup=\sqrt{2}$.}
\label{fig:Ls_dst}
\end{figure*}

\subsection{Sparse Operators of 16 DTTs}
\label{subsec:filter_sparse_all}

\renewcommand{\arraystretch}{1.2}
\begin{table*}[!t]
    \caption{Left and right boundary conditions (b.c.) of 16 DTTs.}
    \label{tab:left_right_bc}
    \centering
    \begin{tabular}{|l|l|c|c|c|c|}
    \hline
         \multicolumn{2}{|c|}{} & \multicolumn{4}{|c|}{Right boundary condition} \\
    \cline{3-6}
         \multicolumn{2}{|c|}{} & \scriptsize$\phi_j(N+k)=\phi_j(N-k)$ & \scriptsize$\phi_j(N+k)=-\phi_j(N-k)$ & \scriptsize$\phi_j(N+k)=\phi_j(N-k+1)$ & \scriptsize$\phi_j(N+k)=-\phi_j(N-k+1)$ \\
    \hline
         \multirow{4}{*}{Left b.c.} & \scriptsize$\phi_j(k)=\phi_j(-k+2)$ & DCT-I & DCT-III & DCT-V & DCT-VII \\
    \cline{2-6}
         & \scriptsize$\phi_j(k)=-\phi_j(-k)$ & DST-III & DST-I & DST-VII & DST-V \\
    \cline{2-6}
         & \scriptsize$\phi_j(k)=\phi_j(-k+1)$ & DCT-VI & DCT-VIII & DCT-II & DCT-IV \\
    \cline{2-6}
         & \scriptsize$\phi_j(k)=-\phi_j(-k+1)$ & DST-VIII & DST-VI & DST-IV & DST-II \\
    \hline
    \end{tabular}
\end{table*}
\renewcommand{\arraystretch}{1}

The approach in Sec.~\ref{subsec:sparse_dct_filters} can be adapted to all 16 DTTs, so that their  corresponding sparse operators can be obtained. In Table~\ref{tab:left_right_bc}, we show left and right boundary conditions of the DTTs. Those properties arise from even and odd symmetries of the cosine and sine functions \cite{puschel2003algebraic}, and can be easily verified based on DTT definitions in Table~\ref{tab:eigenpairs}. As an  illustration, we present in
\if\supplementary1
Appendix~\ref{app:sparse_operators}
\else
\cite{lu2021supplementary,lu2020efficient}
\fi
the derivations for DST-VI, DST-VII, and DCT-V, which share the same right boundary condition with DCT-II, but have different left boundary condition Results for those DTTs with other combinations of left/right boundary condition can be easily extended.

Sparse operators and their associated eigenpairs for all DTTs are listed in Table~\ref{tab:eigenpairs}. Figs.~\ref{fig:Ls_dct} and \ref{fig:Ls_dst} show the operators for $N=6$, which can be easily extended to any arbitrary length. Interestingly, we observe that the non-zero entries in all sparse operators have rectangle-like patterns. Indeed, the 16 DTTs are constructed with combinations of 4 types of left boundary conditions and 4 types of right boundary conditions, associated to 4 types of upper-left rectangle edges and 4 types of lower-right rectangle edges in Figs.~\ref{fig:Ls_dct} and \ref{fig:Ls_dst}, respectively.
We also note that some of the sparse operators in Figs.~\ref{fig:Ls_dct} and \ref{fig:Ls_dst} were already known. Those include  $\Zm_\text{DCT-I}^{(1)}$ \cite{kitajima1980symmetric}, $\Id+\Zm_\text{DCT-III}^{(1)}$ and $\Id+\Zm_\text{DCT-IV}^{(1)}$ \cite{hou1987fast}  (and    \cite{sanchez1995diagonalizing} under a more general framework). In \cite{puschel2003algebraic}, left and right boundary conditions have been exploited to obtain sparse matrices with DTT eigenvectors, which correspond to the first operator $\Zm^{(1)}$ for each DTT. However, to the best of our knowledge, graph operators with $\ell > 1$ (i.e., $\Zm^{(2)}$ to $\Zm^{(N-1)}$ for each DTT) have not been studied in the literature and are introduced here for the first time.

\subsection{Sparse 2D DTT Operators}
\label{subsec:sparse_2d_filters}


\begin{figure*}[t]
  \centering
  \subfigure[]{
  \includegraphics[width=.45\textwidth]{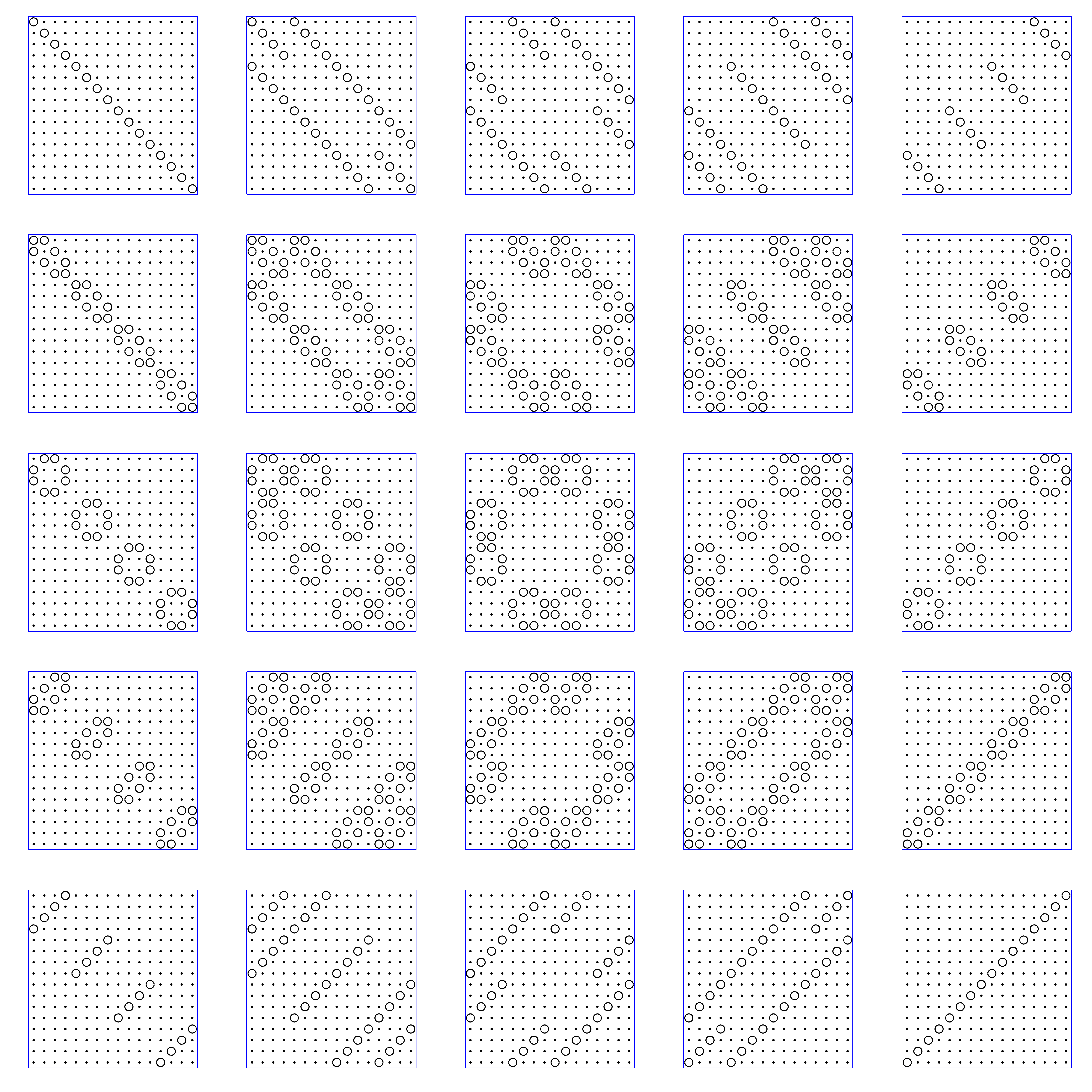}}\hspace{.2cm}
  \subfigure[]{
  \includegraphics[width=.51\textwidth]{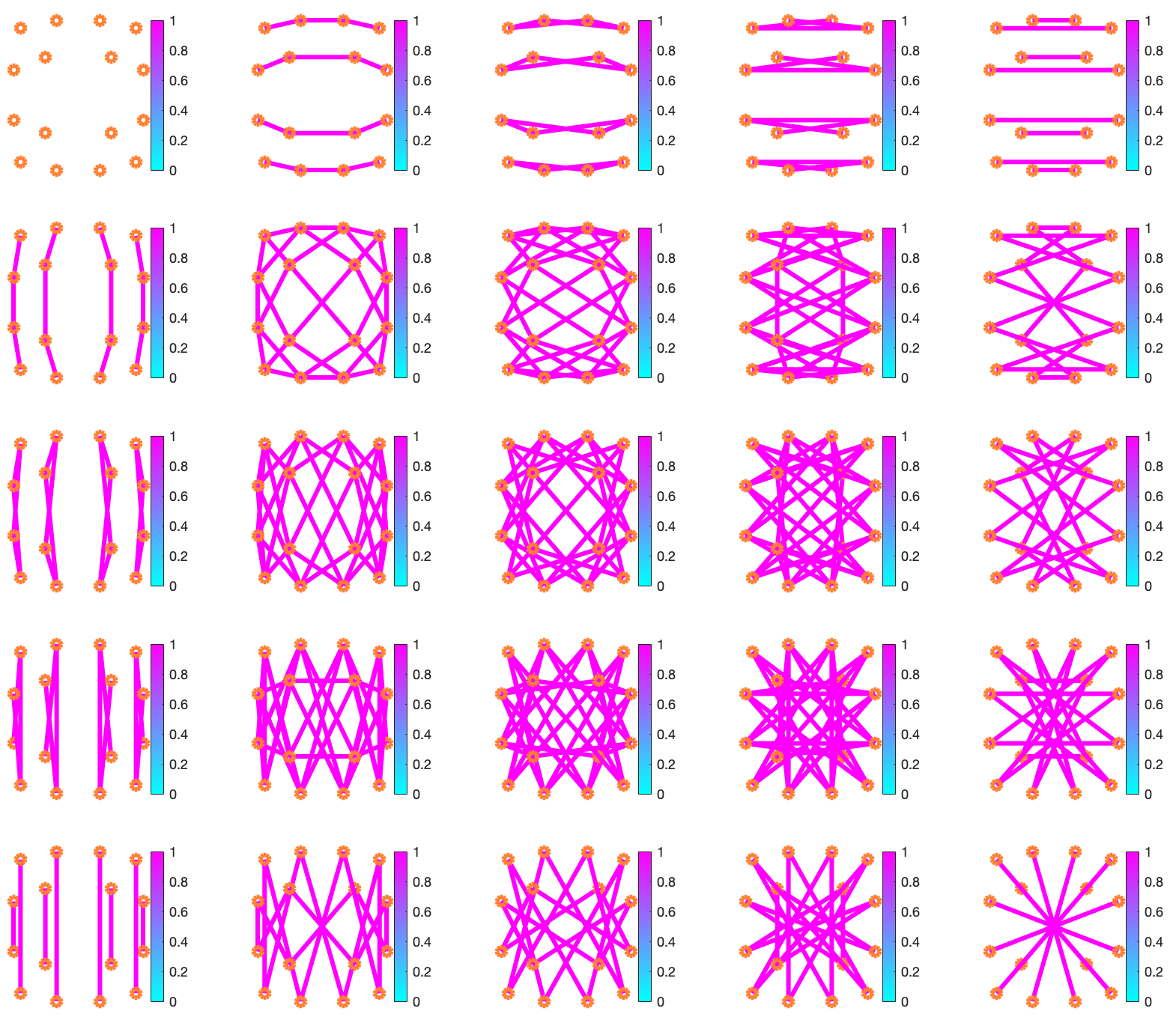}}
  \caption{(a) Sparse operators and (b) graphs associated to 2D $4\times 4$ DCT. Symbols $\cdot$ and $\medcircle$ in (a) represent $0$ and $1$, respectively. For visualization, coordinates in (b) are slightly shifted to prevent some edges from overlapping. Self-loops are not shown in the graphs. The graph in the top-left corner of (b) is associated to the identity matrix, whose corresponding graph contains self-loops only.}
  \label{fig:dct2d}
\end{figure*}

In image and video coding, the DTTs are often applied to 2D pixel blocks, where a combination of 1D DTTs can be applied to columns and rows of the blocks. 
We consider a $N_1\times N_2$ block (with $N_1$ pixel rows and $N_2$ pixel columns),
\[
  \boxed{\begin{array}{llll}
    X_{1,1} & X_{1,2} & \dots & X_{1,N_2} \\
    X_{2,1} & X_{2,2} & \dots & X_{2,N_2} \\
    \multicolumn{1}{c}{\vdots} & \multicolumn{1}{c}{\vdots} & \multicolumn{1}{c}{\vdots} & \multicolumn{1}{c}{\vdots} \\
    X_{N_1,1} & X_{N_1,2} & \dots & X_{N_1,N_2}
  \end{array}}.
\]
We use a 1D vector $\xv\in\mathbb{R}^{N_1 N_2}$ to denote $\Xm$ with column-first ordering:
\[
  \xv=(X_{1,1},X_{2,1},\dots,X_{N_1,1},X_{1,2},X_{2,2},\dots,X_{N1,2},\dots,X_{N_1,N_2})^\top
\]
We assume that the GFT $\Phim=\Phim_r\otimes \Phim_c$ is separable with row transform $\Phim_r$ and column transform $\Phim_c$. 
In such cases, sparse operators of 2D separable GFTs can be obtained from those of 1D transforms:
\begin{proposition}[Sparse 2D DTT operators]
\label{prop:2d}
Let $\Phim=\Phim_r\otimes\Phim_c$ with $\Phim_r$ and $\Phim_c$ being orthogonal transforms among the 16 DTTs, and let $\Zc_r$ and $\Zc_c$ be the set of sparse operators associated to $\Phim_r$ and $\Phim_c$, respectively. Denote the eigenpairs associated to the operators of $\Zc_r$ and $\Zc_c$ as $(\lambda_{r,j},\phiv_{r,j})$ and $(\lambda_{c,k},\phiv_{c,k})$ with $j=1,\dots,N_1$ and $k=1,\dots,N_2$. Then, 
\[ 
  \Zc=\{\Zm_r\otimes\Zm_c,\; \Zm_r\in\Zc_r,\; \Zm_c\in\Zc_c\} 
\]
is a set of sparse operators corresponding to $\Phim_r\otimes\Phim_c$, with associated eigenpairs $(\lambda_{r,j}\lambda_{c,k}, \phiv_{r,j}\otimes\phiv_{c,k})$. 
\end{proposition}
\noindent{\it Proof:}
Let $\Zm_r^{(1)}$, $\dots$, $\Zm_r^{(M_1)}$ be sparse operators in $\Zc_r$ with associated eigenvalues contained in vectors $\lambdav_r^{(1)}$, $\dots$, $\lambdav_r^{(M_1)}$, respectively. Also let $\Zm_c^{(1)}$, $\dots$, $\Zm_c^{(M_1)}$ be those in $\Zc_c$ with eigenvalues in $\lambdav_c^{(1)}$, $\dots$, $\lambdav_c^{(M_2)}$, respectively. We note that
\begin{align*}
  \Zm_r^{(m_1)} &= \Phim_r \cdot\diag(\lambdav_r^{(m_1)}) \cdot\Phim_r^\top, \quad m_1=1,\dots,M_1, \\
  \Zm_c^{(m_2)} &= \Phim_c \cdot\diag(\lambdav_c^{(m_2)}) \cdot\Phim_c^\top, \quad m_2=1,\dots,M_2.
\end{align*}
Applying a well-known Kronecker product identity \cite{zhang2013kronecker}, we obtain
\[
  \pushQED{\qed}
  \Zm_r^{(m_1)}\otimes \Zm_c^{(m_2)} 
  = \Phim \cdot \diag(\lambdav_r^{(m_1)}\otimes\lambdav_c^{(m_2)}) \cdot \Phim^\top.  \qedhere
  \popQED
\]
In Proposition~\ref{prop:2d}, we allow $\Phim_c$ and $\Phim_r$ to be the same. An example is shown in Fig.~\ref{fig:dct2d}, where $\Phim_c=\Phim_r$ is the length-4 DCT-II, and $\Phim$ is the $4\times 4$ 2D DCT.

\subsection{Remarks on Graph Operators of Arbitrary GFTs}
\label{subsec:operators_general}

Obtaining multiple sparse operators $\Zm^{(k)}$ for an arbitrary fixed GFT $\Phim\in\mathbb{R}^{N\times N}$ is a challenging problem in general. Start by noting that given a graph Laplacian associated to $\Phim$ be $\Lm$, with $\lambda_j$ the eigenvalue of $\Lm$ associated to eigenvector $\phiv_j$, if the graph does not have any self-loops, the Laplacian of the complement graph \cite{mohar1991laplacian}
\[
  \Lm^c := N w_\text{max} \Id - w_\text{max}\onev\onev^\top - \Lm,
\]
has eigenpairs $(0,\phiv_1)$ and $(n-\lambda_j, \phiv_j)$ for $j=2,\dots,N$. However, $\Lm^c$ will be a dense matrix when $\Lm$ is sparse, and thus may not be suitable for an efficient MPGF design. 
We next summarize some additional results on the retrieval of sparse graph operators are presented,  
\if\supplementary1
with more details given in Appendix~\ref{app:characterization}. 
\else
with more details given in the supplementary material \cite{lu2021supplementary}.
\fi


\subsubsection{Characterization of Sparse Laplacians of a Common GFT}
Extending a key result in \cite{pasdeloup2018characterization}, we can characterize the set of all graph Laplacians (i.e., those that satisfy \eqref{eq:def_lapl} with non-negative edge and self-loop weights) sharing a given GFT $\Phim$ by a convex polyhedral cone. In particular, those graph Laplacians that are the most sparse among all correspond to the edges of a polyhedral cone (i.e., where the faces of the cone meet each other). However, the enumeration of edges is in general an NP-hard problem since the number of polyhedron vertices or edges can be a combinatorial number of $N$. 

\subsubsection{Construction of Sparse Operators from Symmetric Graphs}
If a graph with Laplacian $\Lm$ satisfies the symmetry property defined in \cite{lu2019fast}, then we can construct a sparse operator in addition to $\Lm$. In particular, we first characterize a node pairing function by an involution $\varphi:\Vc\rightarrow\Vc$, which is a permutation whose inverse is itself (i.e., $\varphi$ satisfies $\varphi(\varphi(i))=i$ for all $i\in\Vc$). In this way, we call a graph $\varphi$-symmetric if $w_{i,j}=w_{\varphi(i),\varphi(j)}$ for all $i,j\in\Vc$. For such a graph, a sparse operator can be constructed as follows:
\begin{lemma}
\label{lem:sym_graph_operator}
Given a $\varphi$-symmetric graph $\Gc$ with Laplacian $\Lm$, we can construct a graph $\overline{\Gc_\varphi}$ by connecting nodes $i$ and $j$ with edge weight 1 for all node pairs $(i,j)$ with $\varphi(i)=j, i\neq j$). In this way, the Laplacian $\overline{\Lm_\varphi}$ of $\overline{\Gc_\varphi}$ commutes with $\Lm$.
\end{lemma}
\if\supplementary1
The proof is presented in Appendix~\ref{app:construction_symmetry}. 
\else
The proof is presented in the supplementary material \cite{lu2021supplementary}. 
\fi


\newcounter{MYtempeqncnt}

\begin{figure*}[!b]
\hrulefill
\setcounter{MYtempeqncnt}{\value{equation}}
\setcounter{equation}{18}  
\begin{equation}
\label{eq:pim}
\scriptsize \Pim_1(\lambdav^{(1)},\dots,\lambdav^{(M)})=
\begin{pmatrix}
1 & \lambda_1^{(1)} & \dots & \lambda_1^{(M)} \\ 
\vdots & \vdots & \vdots & \vdots \\ 
1 & \lambda_N^{(1)} & \dots & \lambda_N^{(M)}
\end{pmatrix}, \quad
\Pim_2(\lambdav^{(1)},\dots,\lambdav^{(M)})=
\begin{pmatrix}
1 & \lambda_1^{(1)} & \dots & \lambda_1^{(M)} & \lambda_1^{(1)}\lambda_1^{(1)} & \lambda_1^{(1)}\lambda_1^{(2)} & \dots & \lambda_1^{(M)}\lambda_1^{(M)} \\ 
\vdots & \vdots & \vdots & \vdots & \vdots & \vdots & \vdots \\ 
1 & \lambda_N^{(1)} & \dots & \lambda_N^{(M)} & \lambda_N^{(1)}\lambda_N^{(1)} & \lambda_N^{(1)}\lambda_N^{(2)} & \dots & \lambda_N^{(M)}\lambda_N^{(M)}
\end{pmatrix}.
\end{equation}
\setcounter{equation}{\value{MYtempeqncnt}}
\end{figure*}

\section{Graph Filter Design with Sparse Operators}
\label{sec:sparse_operators}

\begin{figure}[t]
    \centering
    \includegraphics[width=.45\textwidth]{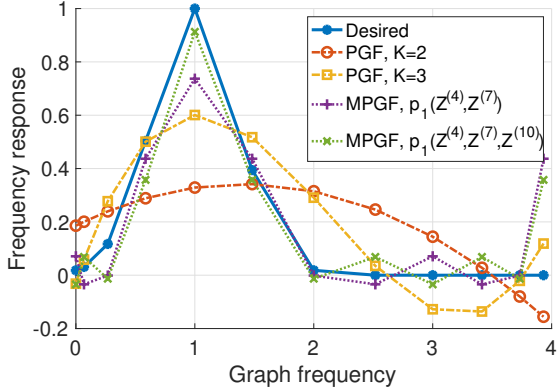}
    \caption{An example for PGF and MPGF fitting results on a length 12 line graph. The desired frequency response is $h^*(\lambda)=\exp(-4(\lambda-1)^2)$. The PGF and MPGF filters have been optimized based on \eqref{eq:pgf} and \eqref{eq:ls_mpgf}.}
    \label{fig:example_mpgf}
\end{figure}

In this section, we introduce some filter design approaches based on sparse operators for DTTs. We start by summarizing the least squares design method  in Section~\ref{subsec:leastsquares}. We also propose a minimax filter design in Section~\ref{subsec:minimax} for both PGF and MPGF. Then, in Section~\ref{subsec:lqf} we show that weighted energy in graph frequency domain can also be efficiently approximated using multiple graph operators.

\subsection{Least Squares (LS) Graph Filter}
\label{subsec:leastsquares}
For an arbitrary graph filter $\Hm^*$, its frequency response, $\hv^*=(h^*(\lambda_1),\dots,h^*(\lambda_N))^\top$, can be approximated with a filter $\Hm_{\Zc,K}$ in \eqref{eq:mpgf} by designing a set of coefficients $\gv$ as in \eqref{eq:mpgf_k1} or \eqref{eq:mpgf_k2}. Let $h(\lambda_j)$ be the frequency response corresponding to $\Hm_{\Zc,K}$, then one way to obtain $\gv$ is through a least squares solution: 
\begin{align}
\label{eq:ls_mpgf}
\gv^* &= \underset{\gv}{\text{argmin}} \quad \sum_{j=1}^N \left( h^*(\lambda_j)-h(\lambda_j)\right)^2 \nonumber\\
&= \underset{\gv}{\text{argmin}} \quad \sum_{j=1}^N \left( h^*(\lambda_j)-p_K(\lambda_j^{(1)},\dots,\lambda_j^{(M)})\right)^2 \nonumber\\
&=\underset{\gv}{\text{argmin}} \quad \left\| \hv^* - \Pim_K(\lambdav^{(1)},\dots,\lambdav^{(M)})\cdot\gv \right\|^2,
\end{align}
where $\Pim_K$ for $K=1$ and $K=2$ are shown in \eqref{eq:pim}.

\setcounter{equation}{\value{MYtempeqncnt}+2}

This formulation can be generalized to a weighted least squares problem, where we allow different weights for different graph frequencies. This enables us to approximate the filter in particular frequencies with higher accuracy. In this case, we consider
\begin{align}
\label{eq:ls_mpgf_weighted}
\gv^* &= \underset{\gv}{\text{argmin}} \quad \sum_{j=1}^N \rho_i^2\left( h^*(\lambda_j)-h(\lambda_j)\right)^2 \nonumber\\
&=\underset{\gv}{\text{argmin}} \quad \left\| \diag(\rhov) (\hv^* - \Pim_K \cdot \gv) \right\|^2,
\end{align}
where $\rho_i\geq 0$ is the weight corresponding to $\lambda_i$. Note that when $\rhov=\onev$, the problem \eqref{eq:ls_mpgf_weighted} reduces to \eqref{eq:ls_mpgf}.

When $\gv$ is sparser, (i.e., its $\ell_0$ norm is smaller), fewer terms will be involved in the polynomial $p_K$, leading to a lower complexity for the filtering operation. This $\ell_0$-constrained problem can be viewed as a sparse representation of $\diag(\rhov)\hv^*$ in an overcomplete dictionary $\diag(\rhov)\Pim_K$. Well-known methods for this problem include the orthogonal matching pursuit (OMP) algorithm \cite{pati1993orthogonal}, and the  optimization with a sparsity-promoting $\ell_1$ constraint:
\begin{equation}
\label{eq:ls_mpgf_l1}
    \underset{\gv}{\text{minimize}} \quad \left\| \diag(\rhov)(\hv^* - \Pim_K \cdot \gv) \right\|^2 \quad
    \text{subject to} \quad \left\| \gv \right\|_1 \leq \tau,
\end{equation}
where $\tau$ is a pre-chosen threshold. In fact, this formulation can be viewed as an extension of its PGF counterpart \cite{shuman2018chebyshev} to an MPGF setting. Note that \eqref{eq:ls_mpgf_l1} is a $\ell_1$-constrained least squares problem (a.k.a., the LASSO problem), where efficient solvers are available \cite{tibshirani1996regression}. 

Compared to conventional PGF $\Hm$ in \eqref{eq:pgf}, the implementation with $\Hm_{\Zc,K}$ has several advantages. First, when $K=1$, the MPGF \eqref{eq:mpgf_k1} is a linear combination of different sparse operators, which is amenable to parallelization. This is in contrast to high degree PGFs based on \eqref{eq:pgf}, which require applying the graph operator repeatedly. Second, $\Hm_{\Zc,K}$ is a generalization of $\Hm$ and provides more degrees of freedom, which provides more accurate approximation with equal or lower order polynomial. Note that, while the eigenvalues of $\Zm^k$ for $k=1,2,\dots$ are typically all increasing (if $\Zm=\Lm$) or decreasing (if, for instance, $\Zm=2\Id-\Lm$), those of different $\Zm^{(m)}$'s have more diverse distributions (i.e., increasing, decreasing, or non-monotonic). Thus, MPGFs provide better approximations for  filters with non-monotonic frequency responses. For example, we demonstrate in Fig.~\ref{fig:example_mpgf} the resulting PGF and MPGF for a bandpass filter. We can see that, for $K=2$ and $K=3$, a degree-1 MPGF with $K$ operators gives a higher approximation accuracy than a degree-$K$ PGF, while they have a similar complexity.


\subsection{Minimax Graph Filter}
\label{subsec:minimax}

The minimax approach is a popular filter design method in classical signal processing. The goal is to design an length-$K$ FIR filter whose frequency response $G(e^{j\omega})$ approximates the desired frequency response $H(e^{j\omega})$ in a way that the maximum error within some range of frequency is minimized. A standard design method is the Parks-McClellan algorithm, which is a variation of the Remez exchange algorithm
\cite{mcclellan1973computer}.

Here, we explore minimax design criteria for graph filters. We denote $h^*(\lambda)$ the desired frequency response, and $g(\lambda)$ the polynomial filter that approximates $h^*(\lambda)$. Source code for the proposed minimax graph filter design methods can be found in \cite{lu2021dct}.

\subsubsection{Polynomial Graph Filter} Let $g(\lambda)$ be the PGF with degree $K$ given by \eqref{eq:pgf}. Since graph frequencies $\lambda_1$, $\dots$, $\lambda_N$ are discrete, we only need to minimize the maximum error between $h^*$ and $g$ at frequencies $\lambda_1$, $\dots$, $\lambda_N$. In particular, we would like to solve polynomial coefficients $g_i$: 
\[
  \underset{\bv}{\text{minimize}} \quad \underbrace{\underset{i}{\text{max}} \; \rho_i \left| h^*(\lambda_i) - \sum_{j=0}^K g_j\lambda_i^j \right|}_{\|\diag(\rhov)\left(\hv^* - \Psim \gv \right)\|_\infty}
\]
where $\Psim$ is the matrix in \eqref{eq:fir_approx_problem}, $\rho_i$ is the weight associated to $\lambda_i$ and $\|\cdot\|_\infty$ represents the infinity norm. Note that, when $K\geq N-1$ and $\Psim$ is full row rank, then $\hv^*=\Psim\gv$ can be achieved with $\gv=\Psim^\dagger\hv^*$. Otherwise, we reduce this problem by setting $\epsilon=\|\diag(\rhov)\left(\hv^*-\Psim\gv\right)\|_\infty$:
\begin{equation}
\label{eq:minimax_pgf}
  \underset{\gv,\;\epsilon}{\text{minimize}} \quad \epsilon 
  \quad \text{subject to} \quad -\epsilon\onev \preceq \diag(\rhov)\left(\hv^*-\Psim\gv\right) \preceq \epsilon\onev,
\end{equation}
whose solution can be efficiently obtained with a linear programming solver.

\subsubsection{Multivariate Polynomial Graph Filter} Now we consider $g(\lambda)$ a graph filter with $M$ graph operators with degree $K$, as in \eqref{eq:mpgf}. In this case, we can simply extend the problem \eqref{eq:minimax_pgf} to 
\begin{equation}
\label{eq:minimax_mpgf}
  \underset{\gv,\;\epsilon}{\text{minimize}} \quad \epsilon 
  \quad \text{subject to} \quad -\epsilon\onev \preceq \diag(\rhov) \left(\hv^*-\Pim_K\gv\right) \preceq \epsilon\onev,
\end{equation}
where a $\ell_1$ or $\ell_0$ norm constraint on $\gv$ can also be considered.

\begin{figure}[t]
    \centering
    \includegraphics[width=.45\textwidth]{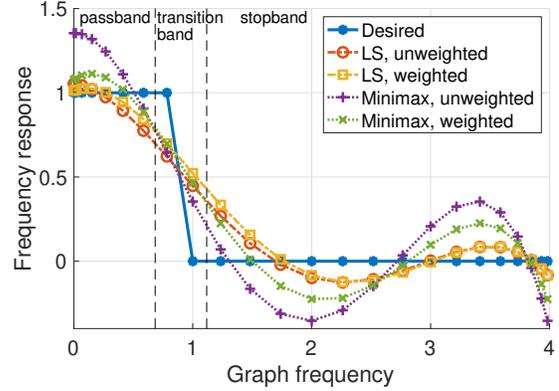}
    \caption{Example illustrating the frequency responses of degree $K=4$ PGF with least squares (LS) and minimax criteria, with weighted or unweighted settings. The filters are defined on a length 24 line graph. In the weighted setting, weights $\rho_i$ are chosen to be 2, 0, and 1 for passband, transition band, and stopband, respectively.}
    \label{fig:example_minimax}
\end{figure}

To summarize, we show in Table~\ref{tab:filter_approaches} the objective functions of least squares and minimax designs with PGF and MPGF, where weights on different graph frequencies are considered. Note that the least squares PGF design shown in Table~\ref{tab:filter_approaches} is a simple extension of the unweighted design \eqref{eq:fir_approx_problem} in \cite{sandryhaila2014discrete}.

Using an ideal low-pass filter as the desired filter, we show a toy example with degree-4 PGF in Fig.~\ref{fig:example_minimax}. When different weights $\rho_i$ are used for passband, transition band, and stopband, approximation accuracies differ for different graph frequencies. By comparing LS and minimax results in a weighted setting, we also see that the minimax criterion yields a smaller maximum error within the passband (see the last frequency bin in passband) and stopband (see the first frequency bin in stopband).

\renewcommand{\arraystretch}{1.5}
\begin{table}[t]
    \centering
    \begin{tabular}{|c|c|c|}
    \hline
         & PGF & MPGF \\
    \hline
        Least squares & $\underset{\gv}{\text{min}} \quad || \diag(\rhov) (\hv - \Psim\gv) ||^2$
        & \eqref{eq:ls_mpgf_weighted} \\
    \hline 
        Minimax & \eqref{eq:minimax_pgf} &  \eqref{eq:minimax_mpgf} \\
    \hline
    \end{tabular}
    \caption{Least squares and minimax design approaches of for PGF and MPGF, with weights $\rho_i$ on different graph frequencies. }
    \label{tab:filter_approaches}
\end{table}
\renewcommand{\arraystretch}{1}

\subsection{Weighted GFT Domain Energy Evaluation}
\label{subsec:lqf}
Let $\xv$ be a signal and $\Phim$ be a GFT to be applied, we consider a weighted sum of squared GFT coefficients:
\begin{equation}
\label{eq:wsqsum}
  \Cc_{\Phim}(\xv;\qv)=\sum_{i=1}^N q_i (\phiv_i^\top\xv)^2,
\end{equation}
where arbitrary weights $\qv=(q_1,\dots,q_N)^\top$ can be considered. Then $\Cc_{\Phim}(\xv;\qv)$ has a similar form to the Laplacian quadratic form \eqref{eq:lqf}, since 
\begin{align}
\label{eq:quadratic_frequency}
  \xv^\top \Lm \xv 
  = \sum_{l=1}^N \lambda_l (\phiv_l^\top\xv)^2.
\end{align}
Note that computation of $\xv^\top\Lm\xv$ using \eqref{eq:lqf} can be done in the vertex domain, and does not require the GFT coefficients. This provides a low complexity implementation than \eqref{eq:quadratic_frequency}, especially when the graph is sparse (i.e., few edges and self-loops).

Similar to vertex domain Laplacian quadratic form computation \eqref{eq:lqf}, we note that $\Cc_{\Phim}(\xv;\qv)$ can also be realized as a quadratic form:
\begin{equation}
\label{eq:wsqsum_filter}
  \Cc_{\Phim}(\xv;\qv)=\sum_{i=1}^N q_i (\phiv_i^\top\xv)^2=\xv^\top \underbrace{\left(\Phim\cdot\diag(\qv)\cdot\Phim^\top\right)}_{\Hm_\qv}\xv,
\end{equation}
where $\Hm_\qv$ can be viewed as a graph filter with frequency response $h_\qv(\lambda_i) = q_i$.
Thus, we can approximate $\Hm_\qv$ with a sparse filter $\Hm_{\hat{\qv}}$ such that $\xv^\top\Hm_{\hat{\qv}}\xv$ approximates $\Cc_{\Phim}(\xv;\qv)$. For example, if we consider a polynomial with degree 1 as in \eqref{eq:mpgf_k1}, we have
\begin{equation}
\label{eq:approx}
  \xv^\top\underbrace{\left[ g_0\Id+\sum_{m=1}^M g_m\Zm^{(m)} \right]}_{\Hm_{\hat{\qv}}}\xv = \sum_{i=1}^N \underbrace{\left(g_0+\sum_{m=1}^M g_m \lambda_i^{(m)}\right)}_{\hat{q_i}} (\phiv_i^\top\xv)^2.
\end{equation}
The left hand side can be computed efficiently if there are only a few nonzero $g_m$, making $\Hm_{\hat{\qv}}$ sparse. The right hand side can be viewed as a proxy of \eqref{eq:wsqsum} if $g_m$'s are chosen such that $\hat{q}_i\approx q_i$. Such coefficients $g_m$ can be obtained by solving \eqref{eq:ls_mpgf_l1} with $\hv^*=\qv$.

\subsection{Complexity Analysis}
\label{subsec:complexity}
For a graph with $N$ nodes and $E$ edges, it has been shown in \cite{coutino2019advances} that a degree-$K$ PGF has $\Oc(KE)$ complexity. For an MPGF with $R$ terms, we denote $E'$ the maximum number of nonzero elements of the operator among all operators involved. Each term of MPGF requires at most $\Oc(KE')$ operations, so the overall complexity of an MPGF is $\Oc(KRE')$. We note that for DTT filters, the sparsity of all operators we have introduced is at most $2N$. Thus, complexities of PGF and MPGF can be reduced to $\Oc(KN)$ and $\Oc(KRN)$, respectively. We note that $\Oc(KRN)$ is not a tight upper bound for the complexity if many terms of the MPGF have lower degrees than $K$. In addition, the polynomial degree required by an MPGF to reach a similar accuracy as a PGF can achieve may be lower. Thus, an MPGF does not necessarily have higher complexity than a PGF that bring a similar approximation accuracy. Indeed, MPGF implementation may be further optimized by parallelizing the computation associated to different graph operators.

\section{Experiments}
\label{sec:experiment}

We consider two experiments to validate the filter design approaches. In Sec.~\ref{subsec:filter_approx}, we evaluate the complexity of PGF and MPGF for DCT-II, and compare the trade-off between complexity and filter approximation accuracy as compared to conventional implementations in the DCT domain. In Sec.~\ref{subsec:pruning} we implement DTT filters in a state-of-the-art video encoder--AV1, where we obtain a computational speedup in transform type search.

\begin{figure*}
\centering
\subfigure[Tikhonov, 16$\times$16 grid]{
\includegraphics[width=.45\textwidth]{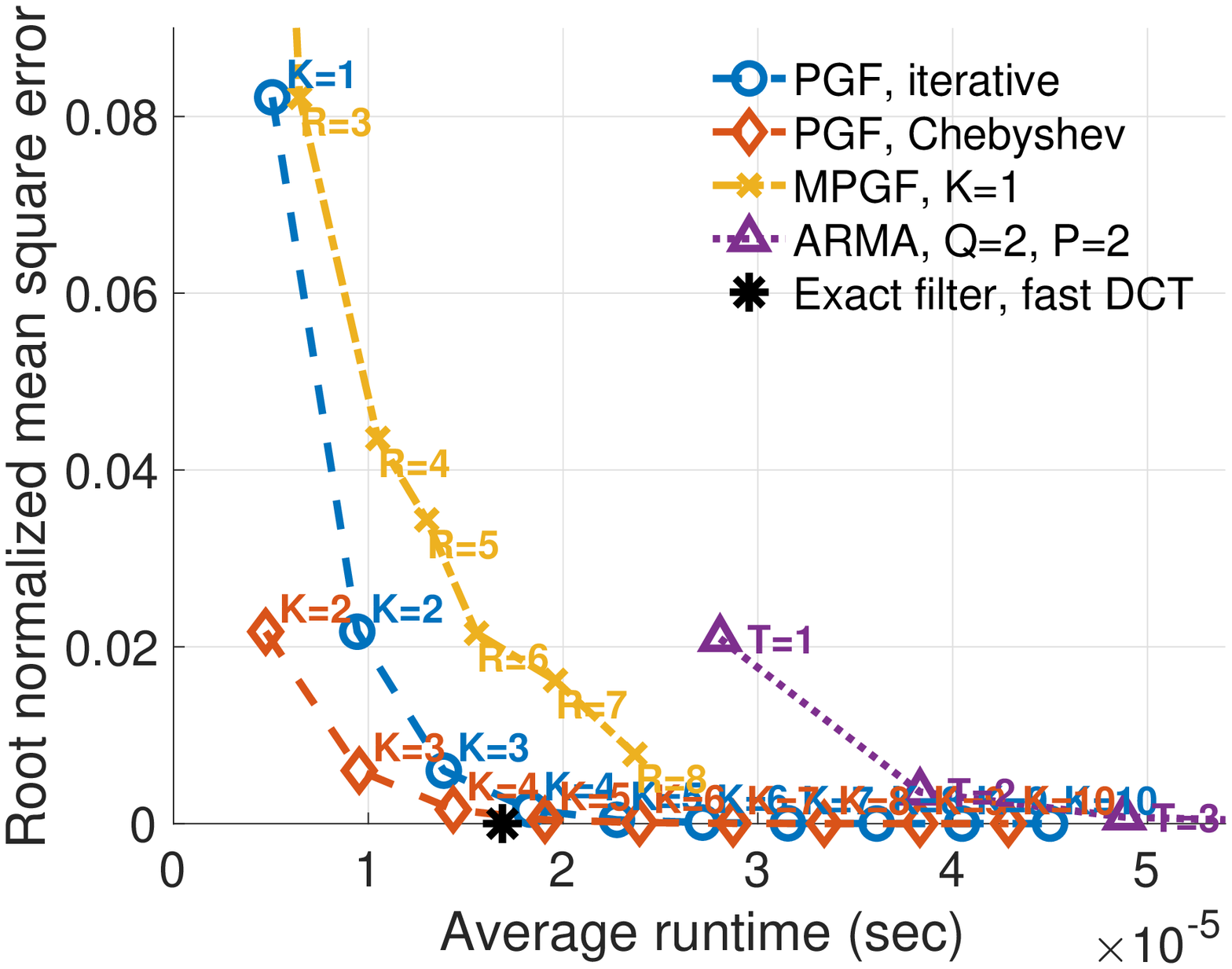}}%
\subfigure[Bandpass exponential, 16$\times$16 grid]{
\includegraphics[width=.45\textwidth]{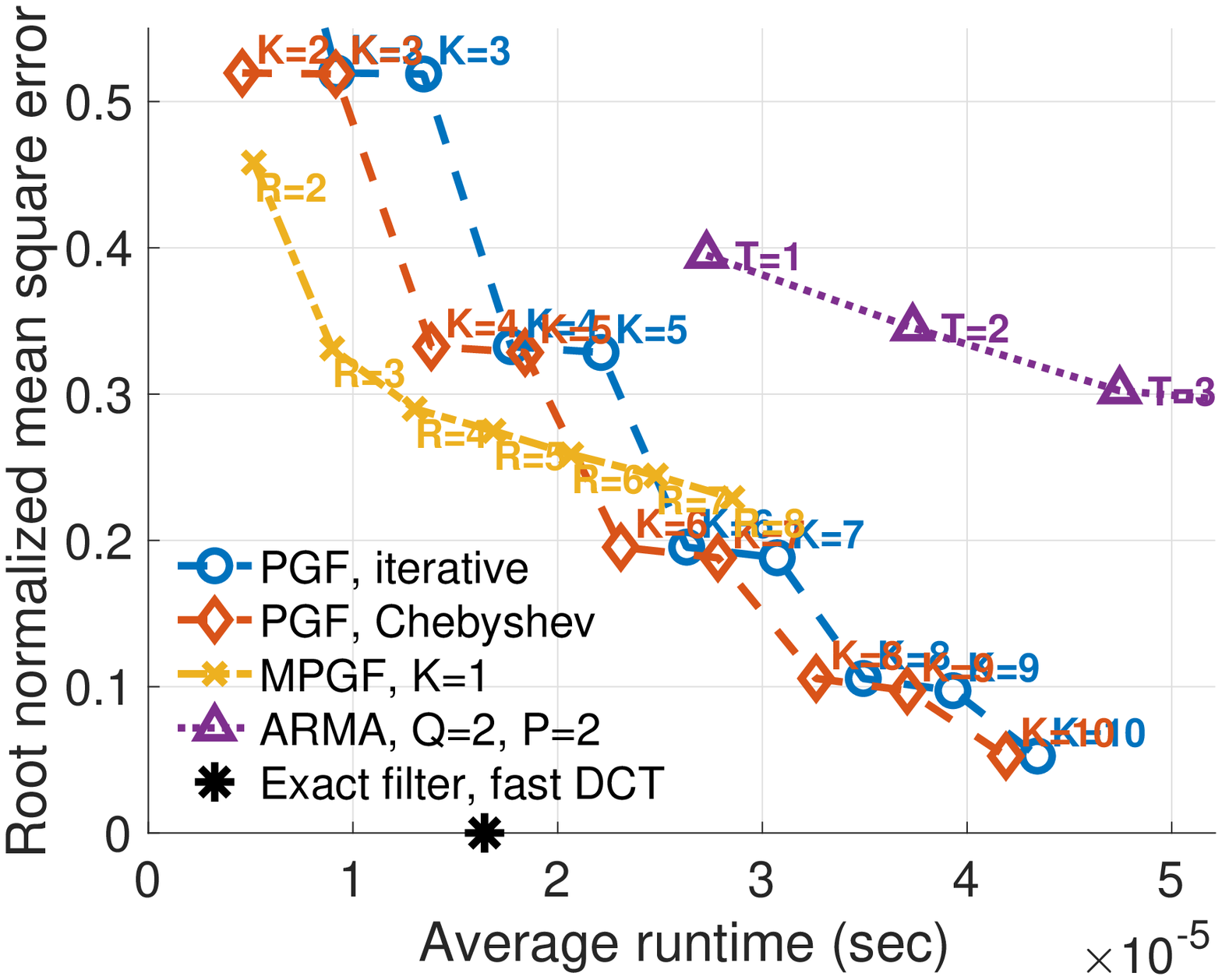}}\\
\subfigure[Tikhonov, length-64 line graph]{
\includegraphics[width=.45\textwidth]{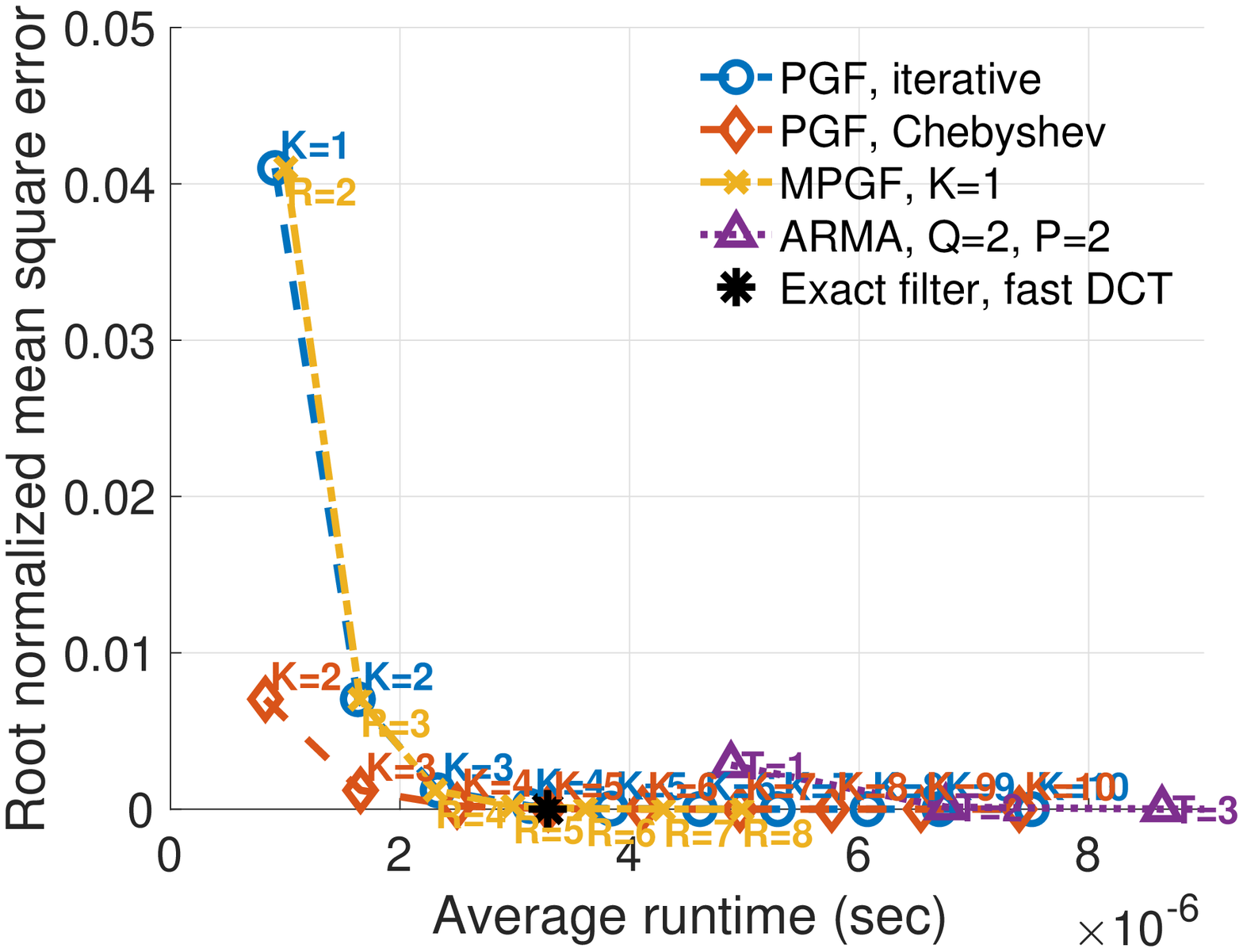}}%
\subfigure[Bandpass exponential, length-64 line graph]{
\includegraphics[width=.45\textwidth]{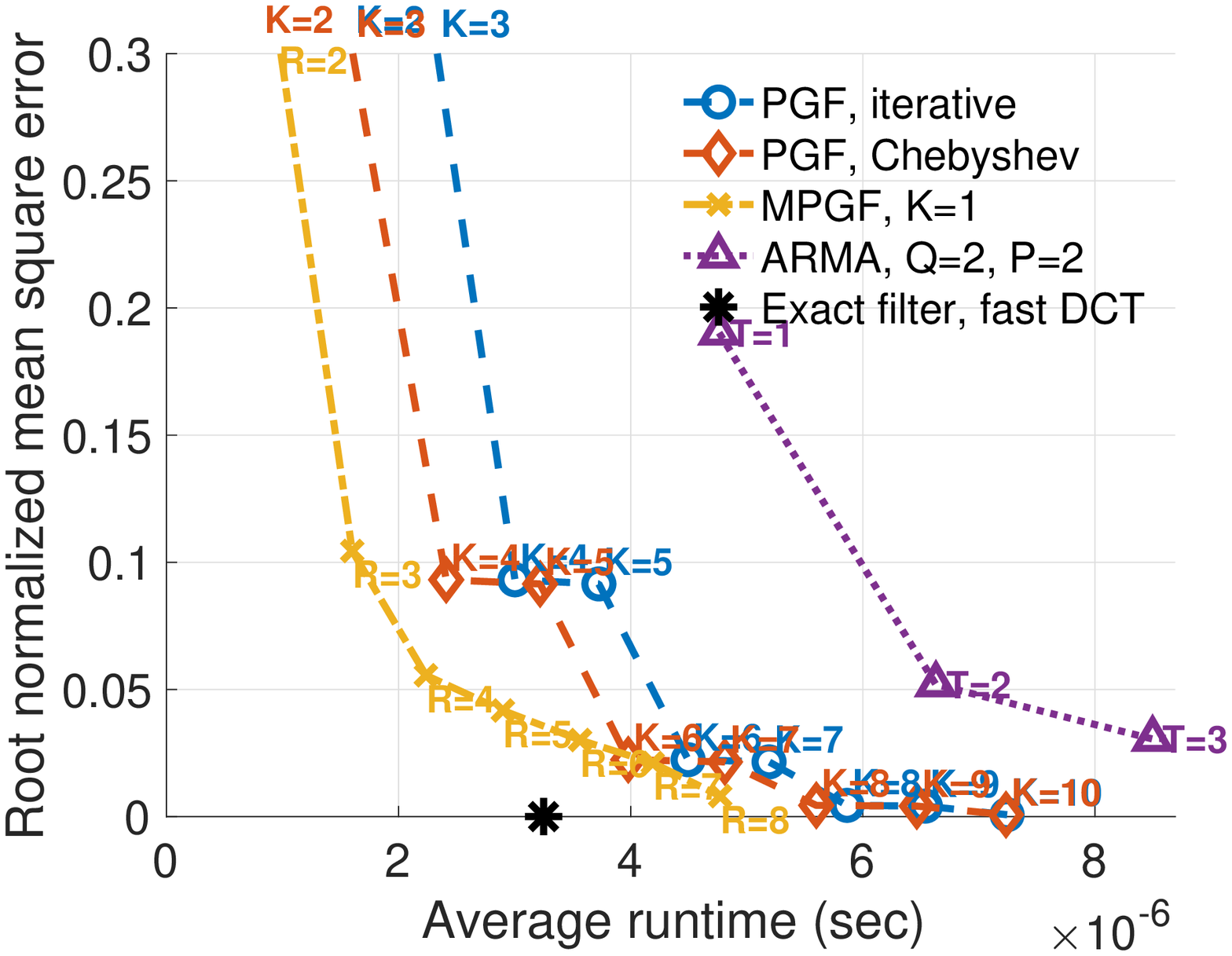}}
\caption{Runtime vs approximation error for (a)(c) Tikhonov DCT filter, (b)(d) bandpass exponential DCT filter. Those filters are defined based on two different graphs: (a)(b) $16\times 16$ grid, (c)(d) length-64 line graph. Different PGF degrees $K$, MPGF operators involved $R$, and ARMA iteration numbers $T$, are labelled in the figures.}
\label{fig:rtvserr}
\end{figure*}

\subsection{Filter Approximation Accuracy with Respect to Complexity}
\label{subsec:filter_approx}

In the first experiment, we implement several DCT filters that have been used in the literature. The graphs we use for this experiment include a $16\times 16$ grid, and a length-64 line graph. Those filters are implemented in C in order to fairly evaluate computational complexity under an environment close to hardware\footnote{The source code for this experiment is available in \cite{lu2021dct}.}. 

\subsubsection{Comparison among filter implementations} Here, the following filters are considered:
\begin{itemize}
    \item \emph{Tikhonov filter:} given $\zv=\xv+\nv$, a noisy observation of signal $\xv$, the denoising problem can be formulated as a regularized least squares problem:
    \begin{align*}
        \underset{\xv}{\text{minimize}} \quad \|\xv-\zv\|^2 + \mu\xv^\top\Lm\xv.
    \end{align*}
    The solution is given by $\hat{\xv}=\Hm_{t}\xv$, where $\Hm_t = (\Id + \mu\Lm)^{-1}$ is known as the Tikhonov graph filter with frequency response $h_{t}(\lambda)=1/(1+\mu\lambda)$. 
    Applications of the Tikhonov filter in graph signal processing include signal denoising \cite{shuman2013emerging}, classification \cite{ma2016diffusion}, and inter-predicted video coding \cite{zhang2015graph}.
    \item \emph{Bandpass exponential filter:} bandpass graph filters are key components in $M$-channel graph filter banks \cite{teke2017extending,tanaka2014m-channel}. Here, we consider the frequency response 
    \[
      h_{\text{exp}}(\lambda)=\exp(-\gamma(\lambda-\lambda_{pb})^2),
    \]
    where $\gamma>0$ and $\lambda_{pb}$ is the central frequency of the passband. 
\end{itemize}
For the choice of parameters, we use $\mu=0.25$, $\gamma=1$, and $\lambda_c=\lambda_{pb}=0.5\lambda_{max}$ in this experiment. The following filter implementations are compared:
\begin{itemize}
    \item \emph{Polynomial DCT filter:} given the desired frequency response, two implementation methods for PGF (with LS design) are considered, namely, \textit{PGFi}, the iterative implementation described in Sec.~\ref{subsec:graph_filter} and  \textit{PGF-C} which implements PGFs using recurrence relations of Chebyshev polynomials \cite{shuman2018chebyshev}. 
    \item \emph{Multivariate polynomial DCT filter:} we consider all sparse graph operators (289 operators for the $16\times 16$ grid and 65 operators for the length-64 line graph). Then, we obtain the least squares filter \eqref{eq:ls_mpgf} with an $\ell_0$ constraint and $K=1$ using orthogonal matching pursuit, with $R$ being 2 to 8.
    \item \emph{Autoregressive moving average (ARMA) graph filter \cite{isufi2017autoregressive}}: we consider an IIR graph filter in rational polynomial form, i.e.,
    \[
      \Hm_{\text{ARMA}}=\left(\sum_{p=0}^Pa_p\Zm^p\right)^{-1}\left(\sum_{q=0}^Qb_q\Zm^q\right).
    \]
    We choose polynomial degrees as $Q=P=2$ and consider different numbers of iterations $T$. The graph filter implementation is based on the conjugate gradient approach described in \cite{liu2019filter}, whose complexity is $\Oc((PT+Q)E)$.
    \item \emph{Exact filter with fast DCT:} the filter operation is performed by a cascade of a forward DCT, a frequency masking with $h$, and an inverse DCT, where the forward and inverse DCTs are implemented using well-known fast algorithms \cite{chen1977fast}. For $4\times 4$ or $16\times 16$ grids, 2D separable DCTs are implemented, where a fast 1D DCT is applied to all rows and columns.
\end{itemize}
In LS designs, uniform weights $\rhov=\onev$ are used. For each graph we consider, 20000 random input signals are generated and  the complexity for each graph filter method is evaluated as an average runtime over all 20000 trials. We measure the error between approximate and exact frequency responses with the root normalized mean square error $\|\hv_{\text{approx}}-\hv\|/\|\hv\|$. 

We show in Fig.~\ref{fig:rtvserr} the resulting runtimes and errors, where a point closer to the origin correspond to a better trade-off between complexity and approximation accuracy. We observe in Fig.~\ref{fig:rtvserr}(a)(c) that low degree PGFs accurately approximate the Tikhonov filter, whose frequency response is closer to a linear function of $\lambda$. In Fig.~\ref{fig:rtvserr}(b)(d), for bandpass exponential filter on the length-64 line graph, MPGF achieves a higher accuracy with lower complexity than PGF and ARMA graph filters. As discussed in Sec.~\ref{subsec:complexity}, the complexity of PGF and MPGF grows linearly with the graph size, while the fast DCT algorithm has $\Oc(N\log N)$ complexity. Thus, PGF and MPGF would achieve a better speed performance with respect to exact filter when the graph size is larger. Note that in this experiment, a fast algorithm with $\Oc(N\log N)$ complexity for the GFT (DCT-II) is available. However, this is not always true for arbitrary graph size $N$, nor for other types of DTTs, where fast exact graph filter may not be available.

\begin{figure}
    \centering
    \subfigure[Ideal low-pass, 16$\times$16 grid]{
    \includegraphics[width=.45\textwidth]{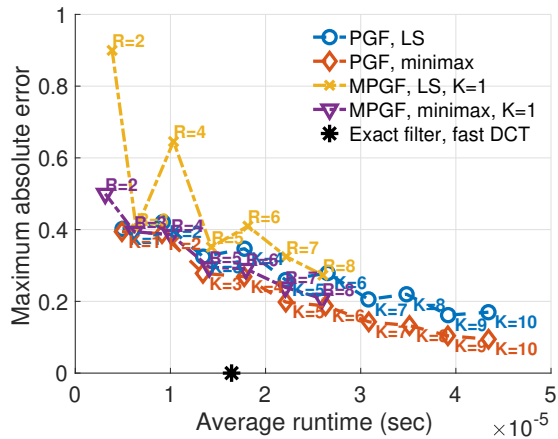}}
    \subfigure[Ideal low-pass, length-64 line graph]{
    \includegraphics[width=.45\textwidth]{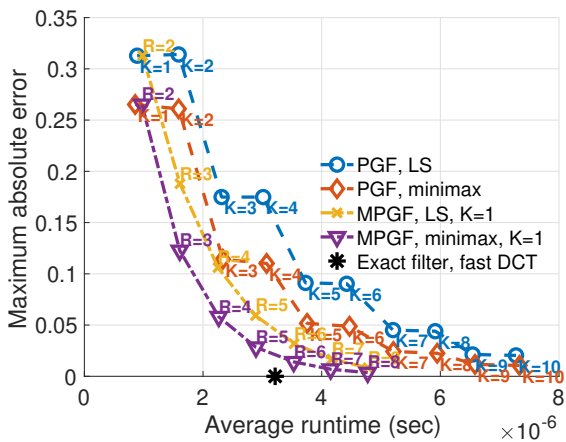}}
    \caption{Runtime vs maximum absolute error for various designs of ideal low-pass filter on (a) 16$\times$16 grid, and (b) length-64 line graph.}
    \label{fig:rtvserr_lp}
\end{figure}

\subsubsection{Evaluation of minimax designs.} 
Next, we consider an ideal low-pass filter:
\[
  h_{LP}(\lambda)=\left\{\begin{array}{ll}1, & 0\leq\lambda\leq\lambda_c \\ 0, & \text{otherwise}\end{array}\right.
\]
where $\lambda_c=0.5\lambda_{max}$ is the cut-off frequency. The weight $\rho_i$ is chosen to be 0 in the transition band $0.4\leq\lambda_i\leq 0.6$, and 1 in passband and stopband. Fig.~\ref{fig:rtvserr_lp} shows the resulting runtimes and approximation errors, which are measured with the maximum absolute error between approximate and desired frequency responses in passband and stopband: $\text{max}_{i}\; \rho_i|h_\text{approx}(\lambda_i)-h(\lambda_i)|$. We can see in Fig.~\ref{fig:rtvserr_lp} that, when $K$ or $R$ increases, the maximum absolute error steadily decreases in PGF and MPGF designs with minimax criteria. In contrast, PGF and MPGF designs with LS criterion may lead to non-monotonic behavior in terms of the maximum absolute error as in Fig.~\ref{fig:rtvserr_lp}(a). In fact, under the LS criterion, using more sparse operators will reduce the least squares error, but does not always decrease the maximum absolute error.

Based on the results in Figs.~\ref{fig:rtvserr} and \ref{fig:rtvserr_lp}, we provide some remarks on the choice of DTT filter implementation:
\begin{itemize}
    \item If the desired frequency response is close to a linear function of $\lambda$, e.g., Tikhonov filters with a small $\mu$ or graph diffusion processes \cite{smola2003kernels}, then a low-order PGF would be sufficiently accurate, and has the lowest complexity.
    \item If the graph size is small, transform length allows a fast DTT algorithm, or when separable DTTs are available (e.g., on a 16$\times$16 grid), DTT filter with fast DTT implementation would be favorable.
    \item For a sufficiently large length (e.g., $N=64$) and a frequency response that is non-smooth (e.g., ideal low-pass filter) or non-monotonic (e.g., bandpass filter), an MPGF design may fit the desired filter with a reasonable speed performance. In particular, we note that $\Zm^{(2})$ is a bandpass filter with passband center $\lambda_{pb}=\lambda_{max}/2$. Thus, MPGF using $\Zm^{(2)}$ would provide an efficiency improvement for bandpass filters with $\lambda_{pb}$ close to $\lambda_{max}/2$.
    \item When robustness of the frequency response in the maximum absolute error sense is an important concern, a design based on minimax criterion would be preferable.
\end{itemize}

\subsection{Transform Type Selection in Video Coding}
\label{subsec:pruning}

In the second experiment, we consider the quadratic form \eqref{eq:wsqsum} as a transform type cost, and apply the method described in Sec.~\ref{subsec:lqf} to speed up transform type selection in the AV1 codec \cite{chen2020overview}. In transform coding \cite{goyal2001theoretical}, \eqref{eq:wsqsum} can be used as a proxy of the bitrate cost for block-wise transform type optimization \cite{hu2015multiresolution,fracastoro2020graph}. In particular, we denote $\xv$ an image or video block, and $\Phim$ the orthogonal transform applied to $\xv$. Lower bitrate cost can be achieved if $\Phim$ gives a high energy compaction in the low frequencies, i.e., the energy of $\Phim^\top\xv$ is concentrated in the first few entries. Thus, the proxy of cost \eqref{eq:wsqsum} can be defined with positive and increasing $\qv$ ($0<q_1<\dots<q_N$) to penalize large and high frequency coefficients, thus favoring transforms having more energy in the low frequencies. 

AV1 includes four 1D transforms: 1)~$\Um$:~DCT, 2)~$\Vm$:~ADST, 3)~$\Jm\Vm$:~FLIPADST, which has flipped ADST functions, and 4)~$\Id$:~IDTX (identity transform), where no transform will be applied. For small inter predicted blocks, all 2D combinations of 1D transforms are used. Namely, there are 16 2D transforms candidates, $(\Tm_\text{col},\Tm_\text{row})$ with $\Tm_\text{col},\Tm_\text{row}\in\{\Um, \Vm,\Jm\Vm,\Id\}$, which makes the encoder computationally expensive. Recent work on encoder complexity reduction includes \cite{su2019machine,lu2018efficient,li2020fast}, which apply heuristic and data-driven techniques to prune transform types during the search.

To speed up transform type selection in AV1, for 1D pixel block $\xv\in\mathbb{R}^N$, we choose the following increasing weights for \eqref{eq:wsqsum}\footnote{As \eqref{eq:wsqsum} is used a proxy of the actual bitrate cost, we leave out the search of optimal weights. Weights are chosen to be increasing functions because transform coefficients associated to a higher frequency typically requires more bits to encode. The weights in \eqref{eq:prune_weights} are used because of their computational for $\Qc_V=\xv_i^\top\Lm_\text{A}\xv_i$. In fact, we have observed experimentally that different choices among several increasing weights produce similar coding results.}:
\begin{equation}
\label{eq:prune_weights}
  q_i=\delta_i=2-2\cos\left(\frac{(i-\frac{1}{2})\pi}{N}\right).
\end{equation}
Then, different transform type costs would be given by \eqref{eq:wsqsum} with different $\Phim$, i.e., $\Cc_{\Tm}(\xv;\qv)$ with $\Tm\in\{\Um, \Vm,\Jm\Vm,\Id\}$. This choice allows efficient computation of exact $\Cc_{\Vm}(\xv;\qv)$ and $\Cc_{\Jm\Vm}(\xv;\qv)$ through their corresponding sparse Laplacian matrices:
\[
  \Cc_{\Vm}(\xv;\qv)=\xv^\top\Lm_\text{A}\xv, \quad 
  \Cc_{\Jm\Vm}(\xv;\qv)=\xv^\top\Jm\Lm_\text{A}\Jm\xv,  
\]
where $\Jm\Lm_\text{A}\Jm$ is the left-right and up-down flipped version of $\Lm_\text{A}$. For the approximation of DCT cost $\Cc_{\Um}(\xv;\qv)$, we obtain $R=3$ nonzeros polynomial coefficients $g_m$ with degree $L=1$ as in \eqref{eq:approx} using an exhaustive search. As a result, costs for all 1D transforms can be computed in the pixel domain as follows
\begin{align}
\label{eq:tx_costs}
 & \Qc_{\Um}=\xv_i^\top \left(g_0\Id+\sum_{m=1}^M g_m \Zm_\text{DCT-II}^{(m)}\right) \xv_i \nonumber\\
 & \Qc_{\Vm}=\Cc_{\Vm}(\xv_i;\qv) = \xv_i^\top\Lm_\text{A}\xv_i  \nonumber\\
 & \Qc_{\Jm\Vm}=\Cc_{\Jm\Vm}(\xv_i;\qv) = \xv_i^\top\Jm\Lm_\text{A}\Jm\xv_i  \nonumber\\
 & \Qc_{\Id}=\Cc_{\Id}(\xv_i;\qv) = \sum_j w_j \xv_i(j)^2,
\end{align}
where $M$ is the number of DCT operators and $g_m$ has only $R=3$ non-zero elements. 

Extending our previous experiment PRUNE\_LAPLACIAN in \cite{lu2018efficient}, we implemented a new experiment named PRUNE\_OPERATORS in AV1\footnote{The experiment has been implemented on a version in July 2020. Available: \url{https://aomedia-review.googlesource.com/c/aom/+/113461}}. We implement the integer versions of the transform cost evaluation \eqref{eq:tx_costs} for transform lengths 4, 8, 16, and 32. Within each 2D block, we take an average over all columns or rows, to obtain column and row costs $\Qc_\Tm^\text{(col)}$ and $\Qc_\Tm^\text{(row)}$ with $\Tm\in\{\Um,\Vm,\Jm\Vm,\Id\}$. Those costs are aggregated into 16 2D transform costs by summing the associated column and row costs. For example, the cost associated to vertical ADST and horizontal DCT is given by 
\[ \Qc_{(\Vm,\Um)}=\Qc_\Vm^\text{(col)}+\Qc_\Um^\text{(row)}. \]

\begin{table}
\centering
\caption{Encoding time and quality loss (in BD rate) of different transform pruning methods. The baseline is AV1 with a full transform search (no pruning). A smaller loss is better.}
\label{tab:pruning}
\begin{tabular}{|c|c|c|}
 \hline
 Method & Encoding time & Quality loss \\
 \hline
 PRUNE\_LAPLACIAN \cite{lu2018efficient} & 91.71\% & 0.32\% \\
 \hline
 PRUNE\_OPERATOR & 89.05\% & 0.31\% \\
 \hline
 PRUNE\_2D\_FAST \cite{su2019machine} & 86.78\% & 0.05\% \\
 \hline 
\end{tabular}
\end{table}

Finally, we design a pruning criteria, where each 2D column (or row) transform will be pruned if its associated cost is relatively large compared to the others. 
\begin{enumerate}
\item[C1.] For $\Tm_\text{col},\Tm_\text{row}\in\{\Um,\Vm,\Jm\Vm\}$, prune $(\Tm_\text{col},\Tm_\text{row})$ if
\begin{align*}
  & \Qc_{(\Tm_\text{col},\Tm_\text{row})}> \\
  & \quad \tau_1\left(\Qc_\Um^\text{(col)}+\Qc_\Vm^\text{(col)}+\Qc_{\Jm\Vm}^\text{(col)}+\Qc_\Um^\text{(row)}+\Qc_\Vm^\text{(row)}+\Qc_{\Jm\Vm}^\text{(row)}\right).
\end{align*}
\item[C2.] For $\Tm_\text{col}=\Id$ or $\Tm_\text{row}=\Id$, prune $(\Tm_\text{col},\Tm_\text{row})$ if
\begin{align*}
  & \Qc_{(\Tm_\text{col},\Tm_\text{row})}> 
  \quad \tau_2\left(\Qc_\Um^\text{(col)}+\Qc_\Vm^\text{(col)}+\Qc_{\Jm\Vm}^\text{(col)}+\Qc_{\Id}^\text{(col)} \right. \\ & \hspace{2.5cm} \left. +\Qc_\Um^\text{(row)}+\Qc_\Vm^\text{(row)}+\Qc_{\Jm\Vm}^\text{(row)}+\Qc_{\Id}^\text{(row)}\right).
\end{align*}
\end{enumerate}
where threshold parameters are chosen as $\tau_1=0.34$ ,$\tau_2=0.33$. Note that the number of 1D transforms being pruned can be different for different blocks. The pruning rules C1 do not depend on $\Qc_\Id$ because IDTX tends to have a larger bitrate cost with a significantly lower computational complexity than the other transforms. Thus, more aggressive pruning criteria C1 is applied to $\Um$, $\Vm$, and $\Jm\Vm$ to reduce more encoding time. 

\begin{table}
\centering
\caption{Encoding time and quality loss (in BD rate) of PRUNE\_OPERATORS versus PRUNE\_2D\_FAST. Smaller or negative loss is better. }
\label{tab:rd_sequences}
\begin{tabular}{|c|c|c|}
 \hline
 Sequence & Encoding time & Quality loss \\
 \hline
 \texttt{akiyo} & 102.10\% & 0.00\% \\
 \texttt{bowing} & 97.22\% & -0.14\% \\
 \texttt{bus} & 103.92\% & -0.17\% \\
 \texttt{city} & 102.36\% & 0.18\% \\
 \texttt{crew} & 103.65\% & 0.07\% \\
 \texttt{foreman} & 104.29\% & 0.07\% \\
 \texttt{harbour} & 106.49\% & -0.06\% \\
 \texttt{ice} & 105.22\% & 0.30\% \\
 \texttt{mobile} & 103.27\% & 0.23\% \\
 \texttt{news} & 103.29\% & -0.09\% \\
 \texttt{pamphlet} & 97.75\% & 0.21\% \\
 \texttt{paris} & 105.54\% & 0.21\% \\
 \texttt{soccer} & 104.53\% & 0.22\% \\
 \texttt{students} & 100.71\% & 0.03\% \\
 \texttt{waterfall} & 102.34\% & 0.23\% \\
 \hline
 Overall & 102.61\% & 0.26\% \\
 \hline 
\end{tabular}
\end{table}

This pruning scheme is evaluated using 15 benchmark test sequences: \texttt{akiyo}, \texttt{bowing}, \texttt{bus}, \texttt{city}, \texttt{crew}, \texttt{foreman}, \texttt{harbour}, \texttt{ice}, \texttt{mobile}, \texttt{news}, \texttt{pamphlet}, \texttt{paris}, \texttt{soccer}, \texttt{students}, and \texttt{waterfall}. The results are shown in Table \ref{tab:pruning}, where the speed improvement is measured in the percentage of encoding time compared to the scheme without any pruning. Each number in the table is an average over several target bitrate levels: 300, 600, 1000, 1500, 2000, 2500, and 3000 kbps. Note that the proposed method yields a smaller quality loss with shorter encoding time than in our previous work \cite{lu2018efficient}. Our method does not outperform the state-of-the-art methods PRUNE\_2D\_FAST in terms of the average BD rate, but shows a gain in particular video sequences such as \texttt{bowing} (as shown in Table~\ref{tab:rd_sequences}. Note that in \cite{su2019machine}, for each supported block size ($N\times N$, $N\times 2N$ and $2N\times N$, with $N\in\{4,8,16\}$), a specific neural network is required to obtain the scores, involving more than 5000 parameters to be learned in total. In contrast, our approach only requires the weights $\qv$ to be determined for each transform length, requiring $4+8+16+32=60$ parameters. With or without optimized weights, our model is more interpretable than the neural-network-based model, as has a significantly smaller number of parameters, whose meaning can be readily explained.

\section{Conclusion}
\label{sec:conclusion}
In this work we explored discrete trigonometric transform (DTT) filtering approaches using sparse graph operators. First, we introduced fundamental graph operators associated to 8 DCTs and 8 DSTs by exploiting trigonometric properties of their transform bases. We also showed that these sparse operators can be extended to 2D separable transforms involving 1D DTTs. 
Considering a weighted setting for frequency response approximation, we proposed least squares and minimax approaches for both polynomial graph filter (PGF) and multivariate polynomial graph filter (MPGF) designs. We demonstrated through an experiment that PGF and MPGF designs would provide a speedup compared to traditional DTT filter implemented in transform domain. We also used MPGF to design a speedup technique for transform type selection in a video encoder, where a significant complexity reduction can be obtained.

\if\supplementary1

\appendices

\section{}
\label{app:sparse_operators}
This appendix presents brief derivations for sparse operators of DST-IV, DST-VII and DCT-V. 

\subsection{Sparse DST-IV Operators}
\label{subsec:filter_sparse_dst4}

Recall the definition of DST-IV functions as in \eqref{eq:dst4}:
\[
  \phi_j(k)=v_j(k)=\sqrt{\frac{2}{N}}\sin \frac{(j-\frac{1}{2})(k-\frac{1}{2})\pi}{N}
\]
As in Section~\ref{subsec:sparse_dct_filters}, we can obtain
\begin{align*}
  & v_j(p-\ell)+v_j(p+\ell) \\
  &=\sqrt{\frac{2}{N}} \left[ \sin \frac{(j-\frac{1}{2})(p-\ell-\frac{1}{2})\pi}{N}+\sin \frac{(j-\frac{1}{2})(p-\ell-\frac{1}{2})\pi}{N} \right] \nonumber\\
  &=2\sqrt{\frac{2}{N}}\sin\frac{(j-\frac{1}{2})(p-\ell-\frac{1}{2})\pi}{N}\cos\frac{\ell(j-\frac{1}{2})\pi}{N} \nonumber\\
  &= \left(2\cos\frac{\ell(j-\frac{1}{2})\pi}{N}\right) v_j(p),
\end{align*}
where we have applied the trigonometric identity
\begin{equation}
\label{eq:sum2prod_sine}
  \sin\alpha+\sin\beta=2\sin\left(\frac{\alpha+\beta}{2}\right)
  \cos\left(\frac{\alpha-\beta}{2}\right).
\end{equation}
By the left and right boundary condition of DST-IV, we have
\begin{align*}
    v_j(p-\ell)=-v_j(-p+\ell+1), \quad
    v_j(p+\ell)=v_j(-p-\ell+2N+1).
\end{align*}
which gives the following result:

\begin{proposition}
\label{prop:dst4}
For $\ell=1,\dots,N-1$, we define $\Zm_\text{DST-IV}^{(\ell)}$ as a $N\times N$ matrix, whose $p$-th row has only two non-zero elements specified as follows:
\begin{align*}
  & {\left(\Zm_\text{DST-IV}^{(\ell)}\right)}_{p,{q_1}}=\left\{\begin{array}{ll}
  1 \text{ with } q_1=p-\ell, & \text{ if } p-\ell \geq 1\\
  -1 \text{ with } q_1=-p+\ell+1, & \text{ otherwise}
  \end{array}\right., \\
  & {\left(\Zm_\text{DST-IV}^{(\ell)}\right)}_{p,{q_2}}=1, \quad q_2 = \left\{\begin{array}{ll}
  p+\ell, & \text{ if } p+\ell \leq N\\
  -p-\ell+2N+1, & \text{ otherwise}
  \end{array}\right.
\end{align*}
The corresponding eigenvalues are $\lambda_j=2\cos\frac{\ell(j-\frac{1}{2})\pi}{N}$.
\end{proposition}

\subsection{Sparse DST-VII Operators}
Now, we consider the basis function of DST-VII:
\[
  \phi_j(k)= \frac{2}{\sqrt{2N+1}} \sin \frac{\left(j-\frac{1}{2}\right)k\pi}{N+\frac{1}{2}}.
\]
Then, by \eqref{eq:sum2prod_sine} we have
\begin{align}
\label{eq:phi_dst7}
    & \phi_j(p-\ell) + \phi_j(p+\ell) \nonumber\\
    &= \frac{2}{\sqrt{2N+1}} \left[ \sin \frac{\left(j-\frac{1}{2}\right)(p-\ell)\pi}{N+\frac{1}{2}} + \sin \frac{\left(j-\frac{1}{2}\right)(p+\ell)\pi}{N+\frac{1}{2}} \right] \nonumber\\
    &= \frac{2}{\sqrt{2N+1}} 2\sin \frac{\left(j-\frac{1}{2}\right)p\pi}{N+\frac{1}{2}} \cos \frac{\ell\left(j-\frac{1}{2}\right)\pi}{N+\frac{1}{2}} \nonumber\\
    &= \left(2\cos \frac{\ell\left(j-\frac{1}{2}\right)\pi}{N+\frac{1}{2}}\right) \phi_j(p)
\end{align}

The left boundary condition (i.e., $\phi_j(k)=-\phi_j(-k)$) of DST-VII corresponds to $\phi_j(p-\ell)=-\phi_j(-p+\ell)$. Together with the right boundary condition $\phi_j(p+\ell)=\phi_j(-p-\ell+2N+1)$, we have the following proposition. 
\begin{proposition}
\label{prop:dst7}
For $\ell=1,\dots,N-1$, we define $\Zm_\text{DST-VII}^{(\ell)}$ as a $N\times N$ matrix, whose $p$-th row has at most two non-zero elements specified as follows:
\begin{align*}
  & {\left(\Zm_\text{DST-VII}^{(\ell)}\right)}_{p,{q_1}}=\left\{\begin{array}{ll}
  1 \text{ with } q_1=p-\ell, & \text{ if } p>\ell \\
  -1 \text{ with } q_1=-p+\ell, & \text{ if } p<\ell
  \end{array}\right., \\
  & {\left(\Zm_\text{DST-VII}^{(\ell)}\right)}_{p,{q_2}}=1, \quad q_2 = \left\{\begin{array}{ll}
  p+\ell, & \text{ if } p+\ell \leq N\\
  -p-\ell+2N+1, & \text{ otherwise}
  \end{array}\right.
\end{align*}
The corresponding eigenvalues are $\lambda_j=2\cos \frac{\ell\left(j-\frac{1}{2}\right)\pi}{N+\frac{1}{2}}$.
\end{proposition}

In Proposition~\ref{prop:dst7}, note that the $\ell$-th row has only one nonzero element because when $p=\ell$, $\phi_j(p-\ell)=0$, and \eqref{eq:phi_dst7} reduces to 
\[
  \phi_j(p+\ell) = \left(2\cos \frac{\ell\left(j-\frac{1}{2}\right)\pi}{N+\frac{1}{2}}\right) \phi_j(p).
\]

\subsection{Sparse DCT-V Operators}
Here, $\phi_j$ are defined as DCT-V basis functions
\[
  \phi_j(k)= \frac{2}{\sqrt{2N-1}} c_j c_k \sin \frac{(j-1)(k-1)\pi}{N-\frac{1}{2}}.
\]
Note that $c_k=1/\sqrt{2}$ for $k=1$ and 1 otherwise. For the trigonometric identity \eqref{eq:sum2prod_cosine} to be applied, we introduce a scaling factor such that $b_k c_k=1$ for all $k$: 
\[
  b_k=\left\{ \begin{array}{ll} \sqrt{2}, & k=1 \\ 1, & \text{otherwise}
  \end{array}.\right.
\]
In this way, by \eqref{eq:sum2prod_cosine} we have
\begin{align*}
    & b_{p-\ell}\cdot\phi_j(p-\ell) + b_{p+\ell}\cdot\phi_j(p+\ell) \\
    & = \frac{2}{\sqrt{2N-1}} c_j \left[ \cos \frac{(j-1)(p-\ell-1)\pi}{N-\frac{1}{2}} + \cos \frac{(j-1)(p+\ell-1)\pi}{N-\frac{1}{2}} \right]\\
    & = \frac{2}{\sqrt{2N-1}} c_j 2\cos \frac{(j-1)(p-1)\pi}{N-\frac{1}{2}} \cos \frac{\ell(j-1)\pi}{N-\frac{1}{2}} \\
    & = \left(2\cos \frac{\ell(j-1)\pi}{N-\frac{1}{2}}\right) b_p \phi_j(p),
\end{align*}
so this eigenvalue equation can be written as
\begin{equation}
\label{eq:bp_dct5}
  \frac{b_{p-\ell}}{b_p}\cdot\phi_j(p-\ell) + \frac{b_{p+\ell}}{b_p}\cdot\phi_j(p+\ell) = \left(2\cos \frac{\ell(j-1)\pi}{N-\frac{1}{2}}\right) \phi_j(p).
\end{equation}

The left boundary condition of DCT-V corresponds to $\phi_j(p-\ell)=\phi_j(-p+\ell+2)$, and the right boundary condition gives $\phi_j(p+\ell)=\phi_j(-p-\ell+2N+1)$. Thus, \eqref{eq:bp_dct5} yields the following proposition:
\begin{proposition}
\label{prop:dct5}
For $\ell=1,\dots,N-1$, we define $\Zm_\text{DCT-V}^{(\ell)}$ as a $N\times N$ matrix, whose $p$-th row has at most two non-zero elements specified as follows:
\begin{align*}
  & {\left(\Zm_\text{DCT-V}^{(\ell)}\right)}_{p,{q_1}}=\left\{\begin{array}{ll}
  \sqrt{2} \text{ with } q_1=1, & \text{ if } p-\ell = 1 \\
  1 \text{ with } q_1=p-\ell, & \text{ if } p-\ell > 1 \\
  1 \text{ with } q_1=-p+\ell+2, & \text{ if } p-\ell \leq 0, p\neq 1
  \end{array}\right., \\
  & {\left(\Zm_\text{DCT-V}^{(\ell)}\right)}_{p,{q_2}}=\left\{\begin{array}{ll}
  \sqrt{2} \text{ with } q_2=1, & p=1 \\
  1 \text{ with } q_2 = p+\ell, & p\neq 1, p+\ell \leq N \\
  1 \text{ with } q_2 = -p-\ell+2N+1, & \text{ otherwise}
  \end{array}\right.
\end{align*}
The corresponding eigenvalues are $\lambda_j=2\cos \frac{\ell(j-1)\pi}{N-\frac{1}{2}}$.
\end{proposition}

The values of $\sqrt{2}$ in Proposition~\ref{prop:dct5} arise from $b_{p-\ell}/b_p$ and $b_{p+\ell}/b_p$ in the LHS of \eqref{eq:bp_dct5}. In particular, when $p=\ell+1$ we have $b_{p-\ell}/b_p=\sqrt{2}$ and $b_{p+\ell}/b_p=1$, so \eqref{eq:bp_dct5} gives
 \[
  \sqrt{2}\phi_j(p-\ell) + \phi_j(p+\ell)= \left(2\cos \frac{\ell(j-1)\pi}{N-\frac{1}{2}}\right) \phi_j(p).
\]
When $p=1$, $b_{p-\ell}/b_p=b_{p+\ell}/b_p=1/\sqrt{2}$. In addition, by the left boundary condition, $\phi_j(p-\ell)=\phi_j(p+\ell)$, so \eqref{eq:bp_dct5} reduces to
\[
  \sqrt{2}\phi_j(p+\ell) = \left(2\cos \frac{\ell(j-1)\pi}{N-\frac{1}{2}}\right) \phi_j(p),\quad \text{for } p=1.
\]
meaning that the first row of $\Zm_\text{DCT-V}^{(\ell)}$ has one nonzero element only.

\section{}
\label{app:characterization}

This appendix includes remarks on the retrieval of sparse operators for general GFTs beyond DCT and DST.

The characterization of all Laplacians that share a common GFT has been studied in the context of graph topology identification and graph diffusion process inference \cite{pasdeloup2018characterization,segarra2017network,decastro2017reconstructing}. In particular, it has been shown in \cite{pasdeloup2018characterization} that the set of \emph{normalized} Laplacian matrices having a fixed GFT can be characterized by a convex polytope. Following a similar proof, we briefly present the counterpart result for \emph{unnormalized} Laplacian with self-loops allowed: 
\begin{theorem}
\label{thm:polyhedral_cone}
The set of Laplacian matrices with a fixed GFT can be characterized by a convex polyhedral cone in the space of eigenvalues $(\lambda_1,\dots,\lambda_N)$.
\end{theorem}
\noindent{\it Proof:}
For a given GFT $\Phim$, let the eigenvalues $\lambda_j$ of the Laplacian be variables. By definition of the Laplacian \eqref{eq:def_lapl}, we can see that 
\begin{equation}
\label{eq:lapl_given_gft}
\Lm=\Phim\cdot\diag(\lambda_1,\dots,\lambda_N)\cdot\Phim^\top=\sum_{k=1}^N \lambda_k\phiv_k\phiv_k^\top
\end{equation}
is a valid Laplacian matrix if $l_{ij}\leq 0$ (non-negative edge weights), $l_{ii} \geq \sum_{j=1, j\neq i}^N l_{ij}$ (non-negative self-loop weights), and $\lambda_k\geq 0$ for all $k$ (non-negative graph frequencies). With the expression \eqref{eq:lapl_given_gft} we have $l_{ij}=\sum_{k=1}^N \lambda_k \phi_k(i)\phi_k(j)$, and thus the Laplacian conditions can be expressed in terms of $\lambda_j$'s:
\begin{align}
\label{eq:polyhedron_constraints}
    & \sum_{k=1}^N \lambda_k \phi_k(i)\phi_k(j)\leq 0, \quad \text{for } i\neq j, \nonumber\\
    & \sum_{k=1}^N \lambda_k \phi_k(i)^2 \geq \sum_{\substack{j=1 \\ j\neq i}}^N\lambda_k\phi_k(i)\phi_k(j), \quad \text{for } i=1,\dots,N, \nonumber\\
    & \lambda_k \geq 0, \quad k=1,\dots,N.
\end{align}
These constraints on $\lambdav=(\lambda_1,\dots,\lambda_N)^\top$ are all linear, so the feasible set for $\lambdav\in\mathbb{R}^N$ is a convex polyhedron. We denote this polyhedron by $\Pc$, and highlight some properties as follows: 
\begin{itemize}
    \item $\Pc$ is non-empty: it is clear to see that $\lambdav=\onev$ gives $\Lm=\Id$, which is a trivial, but valid, Laplacian.
    \item $\lambdav=\zerov$ is the only vertex of $\Pc$: when $\lambda_j=0$ for all $j$, equality is met for all constraints in \eqref{eq:polyhedron_constraints}. This means that all hyperplanes that define $\Pc$ intersect at a common point $\zerov$, which further implies that $\Pc$ does not have other vertices than $\zerov$.
\end{itemize}
From those facts above, we conclude that $\Pc$ is a non-empty convex polyhedral cone.
\qed

\begin{figure}
    \centering
    \includegraphics[width=.35\textwidth]{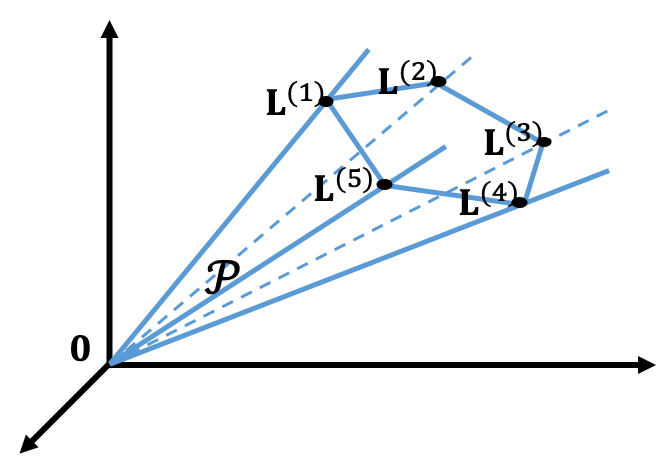}
    \caption{An illustrative example of a polyhedral cone in $\mathbb{R}^3$ with a vertex at $\zerov$ and 5 edges. Any element of the cone can be represented as $\sum_{m=1}^5 a_m \Lm^{(m)}$ with non-negative $a_m$.}
    \label{fig:polyhedral_cone}
\end{figure}

For illustration purpose, we can visualize the structure of a 3-dimensional polyhedral cone with 5 edges in Fig.~\ref{fig:polyhedral_cone}. Notably, any element in $\Pc$ can be expressed by a conical combination (linear combination with non-negative coefficients) of elements on the edges of $\Pc$, as illustrated in Fig.~\ref{fig:polyhedral_cone}. In particular, let $\Pc$ have $M$ edges, and let $\Lm^{(1)}$, $\dots$, $\Lm^{(M)}$ be points on different edges, then any element $\Qm\in\Pc$ can be represented as
\[
  \Qm=\sum_{m=1}^M a_m \Lm^{(m)}, \quad a_m \geq 0.
\]
The fact that Laplacians have non-positive off-diagonal entries implies that the $\Lm^{(m)}$'s are the most sparse Laplacians. This can be seen by noting that a conical combination of two Laplacians must have more non-zero off-diagonal elements than the two individual Laplacians do. 

Since sparse Laplacians are characterized by edges of a polyhedral cone, we can choose sparse operators in \eqref{eq:mpgf} as those Laplacians: $\Zc=\{\Lm^{(k)}\}_k$. The retrieval of those matrices would require an algorithm that enumerates the vertices and edges given the description of a polyhedron. 
A popular algorithm for this problem is the so-called reverse search \cite{avis1992pivoting}, which has a complexity $\Oc(rdv)$, where $r$ is the number of linear constraints in $\mathbb{R}^d$, and $v$ is the number of target vertices. In \eqref{eq:polyhedron_constraints}, $d=N$ and $m=(N^2+3N)/2$, so the complexity reduces to $\Oc(N^3v)$.
In practice, the vertex enumeration problem is in general an NP-hard problem since the number of vertices $v$ can be a combinatorial number: $\binom{r}{d}$. For the purpose of efficient graph filter design, a truncated version of the algorithm \cite{avis1992pivoting} may be applied to obtain a few instead of all vertices. The study of such a truncated algorithm will be left for our future work.

\section{}
\label{app:construction_symmetry}

This appendix shows a construction of sparse operator for graphs with certain symmetry properties. In our recent work \cite{lu2019fast}, we highlighted that a GFT has a butterfly stage for fast implementation if the associated graph demonstrates a symmetry property based on involution permutation (pairing function of nodes):
\begin{definition}
A permutation $\varphi$ on a finite set $\Vc$ is an involution if $\varphi(\varphi(i))=i$, $\forall i\in\Vc$.
\end{definition}
\begin{definition}[\cite{lu2019fast}]
Given an involution $\varphi$ on the vertex set $\Vc$ of graph $\Gc$, then $\Gc$, with a weighted adjacency matrix $\Wm$, is call $\varphi$-symmetric if $w_{i,j}=w_{\varphi(i),\varphi(j)}$, $\forall i\in\Vc, j\in\Vc$.
\end{definition}

With a $\varphi$-symmetric graph $\Gc$, a sparse operator can be constructed as follows.

\begin{lemma}
\label{lem:sym_graph_operator_app}
Given a $\varphi$-symmetric graph $\Gc$ with Laplacian $\Lm$, we can construct a graph $\overline{\Gc_\varphi}$ by connecting nodes $i$ and $j$ with edge weight 1 for all node pairs $(i,j)$ with $\varphi(i)=j, i\neq j$. In this way, the Laplacian $\overline{\Lm_\varphi}$ of $\overline{\Gc_\varphi}$ commutes with $\Lm$.
\end{lemma}
\noindent {\it Proof:} We note that, for $i\in\Vc$, we either have $\varphi(i)=j\neq i$ with $\varphi(j)=i$ or $\varphi(i)=i$. Without loss of generality, we order the graph vertices such that $\varphi(i)=N+1-i$ for $i=1,\dots,k$ and $\varphi(i)=i$ for $i=k+1,\dots,N+1-k$. With this vertex order, we express $\Lm$ in terms of block matrix components,
\[
    \Lm=\begin{pmatrix}
    \Lm_{11} & \Lm_{12} & \Lm_{13} \\
    \Lm_{12}^\top & \Lm_{22} & \Lm_{23} \\
    \Lm_{13}^\top & \Lm_{23}^\top & \Lm_{33}
    \end{pmatrix},
\]
where $\Lm_{11},\Lm_{33}\in\mathbb{R}^{k\times k}$ and $\Lm_{22}\in\mathbb{R}^{(N-2k)\times(N-2k)}$. By $\varphi$-symmetry, the block components of $\Lm$ satisfy (\cite[Lemma 3]{lu2019fast})
\begin{equation}
\label{eq:sym_eqs}
    \Lm_{13}=\Jm\Lm_{13}^\top\Jm, \quad
    \Lm_{33}=\Jm\Lm_{11}\Jm, \quad
    \Lm_{23}=\Lm_{12}^\top\Jm.
\end{equation}

We can also see that the Laplacian constructed from Lemma~\ref{lem:sym_graph_operator_app}, with the same node ordering defined as above, is
\[
    \overline{\Lm_\varphi}=\begin{pmatrix}
    \Id & \zerov & -\Jm \\
    \zerov & \zerov & \zerov \\
    -\Jm & \zerov & \Id
    \end{pmatrix}.
\]
Then, using \eqref{eq:sym_eqs}, we can easily verify that
\[
    \Lm\overline{\Lm_\varphi}=\begin{pmatrix}
    \Lm_{11}-\Lm_{13}\Jm & \zerov & -\Lm_{11}\Jm+\Lm_{13} \\
    \zerov & \zerov & \zerov \\
    \Lm_{13}^\top-\Jm\Lm_{11} & \zerov & -\Lm_{13}^\top\Jm+\Jm\Lm_{11}\Jm
    \end{pmatrix}=\overline{\Lm_\varphi}\Lm,
\]
which concludes the proof.
\qed

\begin{figure}
    \centering
    \subfigure[]{
    \includegraphics[width=.2\textwidth]{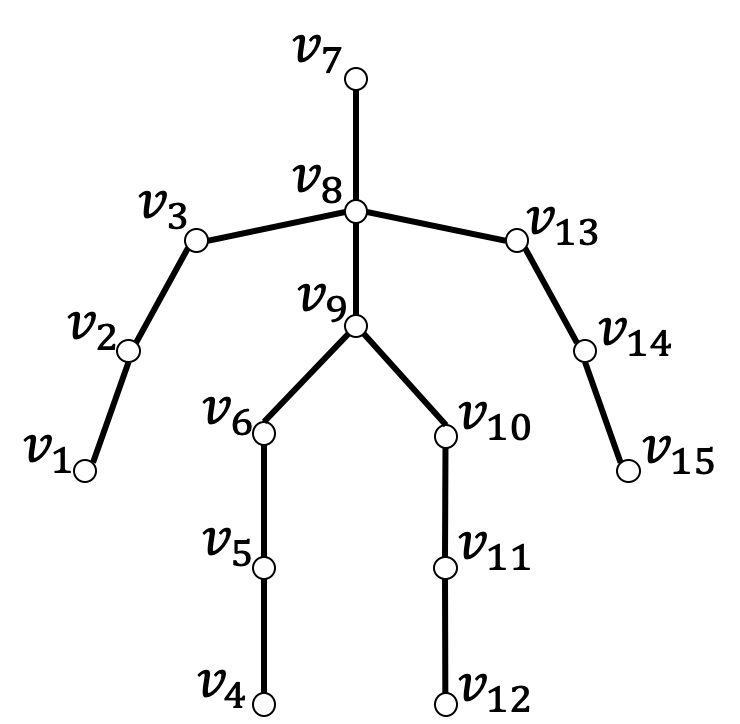}}\hspace{.3cm}
    \subfigure[]{
    \includegraphics[width=.2\textwidth]{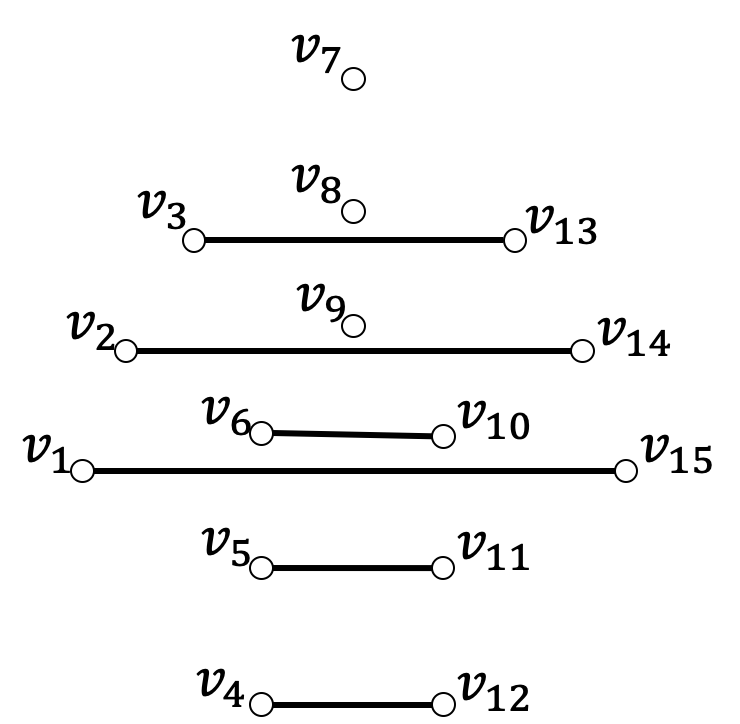}}
    \caption{An illustrative example for graph operator construction based on graph symmetry. (a) The 15-node human skeletal graph $\Gc$. (b) The graph $\overline{\Gc_\varphi}$ associated to an alternative sparse operator by construction. All edge weights are 1.}
    \label{fig:sk15_example}
\end{figure}

We demonstrate an example for the construction of $\overline{\Gc_\varphi}$,  in Fig.~\ref{fig:sk15_example}. Fig.~\ref{fig:sk15_example}(a) shows a 15-node human skeletal graph $\Gc$ \cite{kao2014graph-based}. A left-to-right symmetry can be observed in $\Gc$, which induces an involution $\varphi$ with $\varphi(i)=i$ for $i=7,8,9$ and $\varphi(i)=16-i$ otherwise. With the construction in Lemma~\ref{lem:sym_graph_operator_app}, we obtain a graph $\overline{\Gc_\varphi}$ as in Fig.~\ref{fig:sk15_example}(b) by connecting all pairs of symmetric nodes in Fig.~\ref{fig:sk15_example}(a). We denote $\Zm^{(1)}=\Lm$ and $\Zm^{(2)}=\overline{\Lm_\varphi}$ the Laplacians of $\Gc$ and $\overline{\Gc_\varphi}$, respectively, and $\Psim=(\psiv_1,\dots,\psiv_{15})$ the GFT matrix of $\Lm$ with basis functions in increasing order of eigenvalues. In particular, we have
\begin{align*}
  & \Zm^{(2)}=\Psim\cdot\diag(\lambdav^{(2)})\cdot\Psim^\top, \\ & \lambdav^{(2)}=(0,0,2,2,0,0,0,2,2,0,0,2,2,0,0)^\top.
\end{align*}
Since $\Zm^{(2)}$ has only two distinct eigenvalues with high multiplicities, every polynomial of $\Zm^{(2)}$ also has two distinct eigenvalues only, which poses a limitation for graph filter design. However, an MPGF with both $\Zm^{(1)}$ and $\Zm^{(2)}$ still provides more degrees of freedom compared to a PGF with a single operator.

\fi 

\ifCLASSOPTIONcaptionsoff
  \newpage
\fi

\bibliographystyle{IEEEbib}
\bibliography{refs}







\end{document}

%% file: macros.tex
\usepackage{bm}

\long\def\comment#1{}

\newfont{\bbb}{msbm10 scaled 700}

\newfont{\bb}{msbm10 scaled 1100}

\newcommand{\bv}{{\bf b}}

\newcommand{\gv}{{\bf g}}
\newcommand{\hv}{{\bf h}}

\newcommand{\nv}{{\bf n}}

\newcommand{\qv}{{\bf q}}

\newcommand{\tv}{{\bf t}}
\newcommand{\uv}{{\bf u}}

\newcommand{\vv}{{\bf v}}
\newcommand{\xv}{{\bf x}}
\newcommand{\yv}{{\bf y}}
\newcommand{\zv}{{\bf z}}
\newcommand{\zerov}{{\bf 0}}
\newcommand{\onev}{{\bf 1}}

\newcommand{\Am}{{\bf A}}

\newcommand{\Dm}{{\bf D}}

\newcommand{\Hm}{{\bf H}}
\newcommand{\Id}{{\bf I}}
\newcommand{\Jm}{{\bf J}}

\newcommand{\Lm}{{\bf L}}

\newcommand{\Qm}{{\bf Q}}

\newcommand{\Tm}{{\bf T}}
\newcommand{\Um}{{\bf U}}
\newcommand{\Wm}{{\bf W}}
\newcommand{\Vm}{{\bf V}}
\newcommand{\Xm}{{\bf X}}

\newcommand{\Zm}{{\bf Z}}

\newcommand{\Cc}{{\cal C}}

\newcommand{\Ec}{{\cal E}}

\newcommand{\Gc}{{\cal G}}

\newcommand{\Nc}{{\cal N}}
\newcommand{\Oc}{{\cal O}}
\newcommand{\Pc}{{\cal P}}
\newcommand{\Qc}{{\cal Q}}

\newcommand{\Vc}{{\cal V}}

\newcommand{\Zc}{{\cal Z}}

\newcommand{\Lcb}{{\bm {\mathcal L}}}

\newcommand{\lambdav}{\hbox{\boldmath$\lambda$}}

\newcommand{\phiv}{\hbox{\boldmath$\phi$}}
\newcommand{\psiv}{\hbox{\boldmath$\psi$}}

\newcommand{\rhov}{\hbox{\boldmath$\rho$}}

\newcommand{\Lambdam}{\hbox{\boldmath$\Lambda$}}

\newcommand{\Phim}{\hbox{\boldmath$\Phi$}}
\newcommand{\Pim}{\hbox{\boldmath$\Pi$}}
\newcommand{\Psim}{\hbox{\boldmath$\Psi$}}
\newcommand{\Thetam}{\hbox{\boldmath$\Theta$}}

\newcommand{\diag}{{\hbox{diag}}}

\newtheorem{theorem}{Theorem}

\newtheorem{proposition}{Proposition}
\newtheorem{lemma}{Lemma}
\newtheorem{definition}{Definition}